\documentclass[lettersize,journal]{IEEEtran}
\usepackage{amsmath,amsfonts}
\usepackage{algorithmic}
\usepackage{array}
\usepackage[caption=false,font=normalsize,labelfont=sf,textfont=sf]{subfig}
\usepackage{textcomp}
\usepackage{stfloats}
\usepackage{url}
\usepackage{verbatim}
\usepackage{graphicx}
\usepackage{multirow}
\usepackage{balance}
\usepackage[colorlinks=true, linkcolor=blue, citecolor=green, urlcolor=red]{hyperref} 
\usepackage{xcolor}
\usepackage{algorithm}
\usepackage{algorithmic}
\usepackage{colortbl}
\definecolor{mycolor}{RGB}{0,255,0}

\renewcommand{\arraystretch}{1.0}

\def\BibTeX{{\rm B\kern-.05em{\sc i\kern-.025em b}\kern-.08em
    T\kern-.1667em\lower.7ex\hbox{E}\kern-.125emX}}

\begin{document}
\title{Pseudo-Label Guided Real-World Image De-weathering: \\
A Learning Framework with Imperfect Supervision}

\author{
    Heming Xu$^{1}$, Xiaohui Liu$^{1}$, Zhilu Zhang$^{1}$, Hongzhi Zhang$^{1}$, Xiaohe Wu$^{1,*}$, Wangmeng Zuo$^{1}$
\\
    $^{1}$Harbin Institute of Technology
}

\maketitle

\begin{abstract}
Real-world image de-weathering aims at removing various undesirable weather-related artifacts, \emph{e.g.}, rain, snow, and fog. To this end, acquiring ideal training pairs is crucial.
Existing real-world datasets are typically constructed paired data by extracting clean and degraded images from live streams of landscape scene on the Internet.
Despite the use of strict filtering mechanisms during collection, training pairs inevitably encounter inconsistency in terms of lighting, object position, scene details, etc, making de-weathering models possibly suffer from deformation artifacts under non-ideal supervision.
In this work, we propose a unified solution for real-world image de-weathering with non-ideal supervision, \emph{i.e.}, a pseudo-label guided learning framework, to address various inconsistencies within the real-world paired dataset.
Generally, it consists of a de-weathering model (De-W) and a Consistent Label Constructor (CLC), by which restoration result can be adaptively supervised by original ground-truth image to recover sharp textures while maintaining consistency with the degraded inputs in non-weather content through the supervision of pseudo-labels.
Particularly, a Cross-frame Similarity Aggregation (CSA) module is deployed within CLC to enhance the quality of pseudo-labels by exploring the potential complementary information of multi-frames through graph model.  
Moreover, we introduce an Information Allocation Strategy (IAS) to integrate the original ground-truth images and pseudo-labels, thereby facilitating the joint supervision for the training of de-weathering model.
Extensive experiments demonstrate that our method exhibits significant advantages when trained on imperfectly aligned de-weathering datasets in comparison with other approaches. 
%
%

\end{abstract}

\begin{figure*}[htbp]
    \centering
    \includegraphics[width=0.95\linewidth]{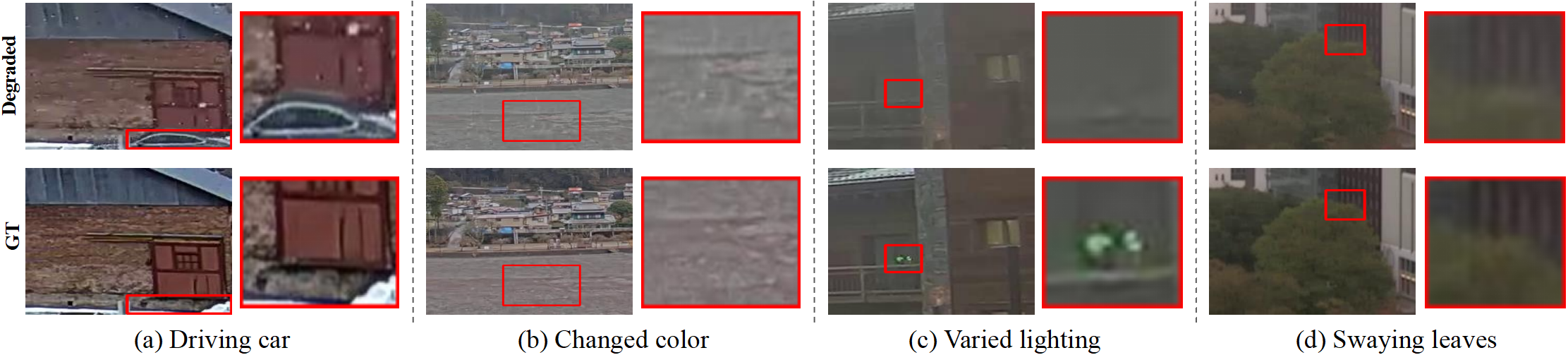} \vspace{-1em}
    \caption{Examples of misalignment categories between the degraded images (top) and the original GT (bottom). These instances are derived from two real-world de-weathering datasets. (\textit{i.e.}, GT-Rain-Snow \cite{ba2022not} and WeatherStream \cite{zhang2023weatherstream}). The pairs in (b) and (c) show inconsistencies in color and lighting. (a) and (d) indicate misalignment caused by object motion, eg., moving vehicle, swaying leaves. The inconsistent contents are highlighted in red boxes.}
    \label{fig:dataset_problems}
\end{figure*}
\section{Introduction} 
\label{sec:1}
\IEEEPARstart{I}{mages} captured under adverse weather conditions, \emph{e.g.}, rain, snow, or fog, exhibit poor perceptual quality in terms of visual fidelity and present significant challenges for vision understanding tasks, such as autonomous driving~\cite{xia2023image,pun2024neural}, depth estimation~\cite{fu2018deep,li2023learning}, detection~\cite{wang2023yolov7,xiong2023cape,tao2023weakly}, segmentation~\cite{zhao2019multi}, etc. 
Image de-weathering, which aims at removing undesirable weather-related artifacts is significant in improving the performance of those tasks.
In recent years, the significant advancements in deep learning have propelled the development of data-driven learning-based de-weathering methods~\cite{li2020all,chen2022learning,fu2017clearing}. 
However, these state-of-the-art techniques typically rely on a substantial number of training pairs.
In practice, obtaining ideal real-world pairs for the purpose of training networks presents a substantial difficulty, as it is virtually impossible to simultaneously acquire images of the same scene with and without weather-related artifacts at the same time.
An initially common choice is to construct paired data by simulating weather degradation for supervised learning. 
Nevertheless, the synthesized generated degraded images frequently display inadequate realism in representing sophisticated and fluctuating nature of real-world weather degradations, like the shapes of raindrops and the density of fog. As a result, models trained on these synthetic pairs commonly face difficulties in accurately generalizing to authentic severe weather, potentially leading to the emergence of artifacts and the forfeiture of original image details.
Alternatively, some studies focus on developing unsupervised~\cite{jin2019unsupervised,chang2023unsupervised} or self-supervised~\cite{liang2022self,bae2022slide} algorithms to eliminate the necessity for paired data during training. Although some of these methods have shown promising results, challenges persist in effectively disentangling the intricate overlaps between foreground and background information in real-world degraded images.
To facilitate the extension of de-weathering models to real-world images, recent research has made substantial progress in the construction of the real paired training datasets.
Given the impracticality of obtaining pairs of clean and weather degraded images at the same timestamp, these methods relax the stringent demand for concurrent capture of ideal image pairs, opting instead to extract paired degraded and clean images from real-time streams of landscape scenes.
For example, GT-RAIN~\cite{ba2022not} dataset introduces time multiplexed pairs, which are obtained by extracting adjacent frames from the video sequences collected from YouTube live streams, encompassing scenes with and without weather-related degradation.
This approach is only effective when nearby frames are captured at the appropriate moment, specifically under ideal scene conditions, such as consistent illumination, and minimal dynamic components.
Moreover, such data collection scheme extremely requires human annotation, making it a labor-intensive task.
As an extension, WeatherStream~\cite{zhang2023weatherstream} further utilizes light transport principles to automatically select time-multiplexed pairs, thus efficiently constructing a large-scale all-weather removal dataset.
And the conclusion indicates that these time multiplexed pairs bridge the domain gap better than the synthetic and semi-real data, which has significant value for the advancement of real-world weather restoration. 
However, it should be noted that, despite the adoption of stringent collection criteria during the dataset collection process to ensure spatial alignment and illumination consistency between the ground truth (GT) and degraded images, certain inconsistencies remain, such as suddenly appearing vehicles, swaying leaves, or changes in lighting and color, as shown in Fig. \ref{fig:dataset_problems}, which inevitably impact the learning process of the de-weathering model negatively.

\begin{figure}[ht]
    \centering
    \includegraphics[width=0.95\linewidth]{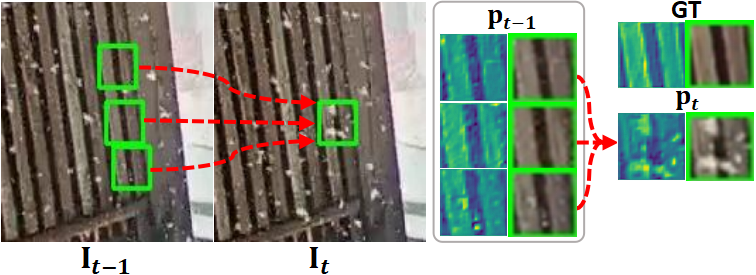}
    \caption{For the query patch $\mathbf{p}_{t}$ in the current frame $\mathbf{I}_{t}$ being processed, its matched patches $\mathbf{p}_{t-1}$ in the neighboring frame $\mathbf{I}_{t-1}$ that have similar textures to the query patch may not be obscured by snowflakes. Each patch's left image represents the feature map obtained from feature extractor VGG19. $\mathbf{p}_{t-1}$ exhibits smaller feature-space distances to the ground truth and can serve as supplementary information.}
    \label{fig:adj_patch}
\end{figure}
In this paper, we propose a unified solution addressing the various inconsistencies within the real-world dataset, \emph{i.e.}, a pseudo-label guided learning framework, which enables training the de-weathering model with imperfect supervision.
Specifically, we design a Consistent Label Constructor (CLC) for generating pseudo-labels to ensure consistency with the degraded inputs in non-weather content. 
Taking the advantages of the dataset derived from videos, we further deploy a Cross-frame Similarity Aggregation (CSA) module to integrate multi-frame information to enhance the quality of pseudo-labels.
Inspired by the theory of image self-similarity~\cite{zontak2011internal}, we extend this concept to sequence frames, as shown in Fig.~\ref{fig:adj_patch}. For each query patch in the degraded input feature space, we match cross-frame self-similar patches in the corresponding regions of neighboring frames. Note that this figure is only illustrative; the actual matching process is achieved with very tiny patch within feature space. These matched patches serve as query-child nodes, forming a directed graph structure that captures their feature aggregation relationships. A Graph Attention Network (GAT)~\cite{velivckovic2017graph} is then adopted for non-local feature aggregation, enhancing the representation capability. 
The enhanced features contribute to obtaining higher-quality pseudo-labels and are subsequently utilized to distill the features of the restoration network, thereby improving the restoration results.
Given that the produced pseudo-labels might not yet achieve the same level of degradation removal as the original labels, we then introduce the Information Allocation Strategy (IAS) to integrate the original and pseudo-labels, thereby facilitating the joint supervision for the de-weathering model.

This work is an extension of the conference paper previously presented in AAAI2024 \cite{liu2024learning}, upon which this manuscript has made three major improvements: 
(\emph{i}) We deploy Cross-frame Similarity Aggregation (CSA) in the original Consistent Label Constructor (CLC) to improve the quality of generated pseudo-labels, as well as the feature representation ability. 
(\emph{ii}) We incorporate feature distillation into the learning of the de-weathering model, aiming to distill the fused features learned from multiple frames to enhance the restoration results.
(\emph{iii}) We conduct additional comparative and ablation experiments to validate the performance of our enhanced pipeline and the effectiveness of each improvement. 
(\emph{iv}) We provide a broader review of the de-weathering literature, discussing the limitations of various approaches proposing a unified solution for real-world de-weathering tasks under imperfect supervision, and reducing the demands on real-world dataset.

Our contributions can be summarized as follows:
\begin{itemize}
\item We introduce a unified solution addressing the various inconsistency issues in real-world datasets, \emph{i.e.}, a pseudo-label guided learning framework, which enables training the de-weathering model with non-ideal supervision.
\item We design a Consistent Label Constructor (CLC) equipped with a Cross-frame Similarity Aggregation (CSA) module to generate high-quality pseudo-labels from multiple frames that maintain consistency with the degraded input images in non-weather content. 
\item We combine the pseudo-labels with the original labels to jointly supervise the learning of de-weathering model through the Information Allocation Strategy (IAS).
\item  Extensive experiments prove that our method outperforms various state-of-the-art methods when trained on real datasets with imperfectly aligned labels.
\end{itemize}

\section{Related Work} \label{sec:2}
\subsection{De-weathering Methods} \label{sec:2.1}
\subsubsection{Single Weather Types}
In recent years, the field of image restoration under adverse weather conditions has witnessed significant advancements, with a multitude of outstanding research outcomes emerging. These methods leverage either traditional approaches or deep learning techniques to address image degradation caused by weather conditions, focusing primarily on three key areas: deraining \cite{oord2018representation,yang2017deep,fu2017clearing,zhang2018density,hu2019depth,chen2023learning, wei2023raindiffusion}, desnowing \cite{liu2018desnownet,li2019single,chen2021all,kaihao2021deep,zhang2021deep,chen2023msp}, and dehazing \cite{he2010single,li2018benchmarking,li1707all,engin2018cycle,qin2020ffa,guo2022image,yang2022self,wu2023ridcp}. By employing advanced technologies such as, traditional image processing, convolutional networks, Generative Adversarial Networks (GANs), Transformers, and diffusion models, these approaches have greatly advanced the progress of image restoration under single weather conditions.
\subsubsection{Multiple Weather Types}
Although advancements have been achieved with those approaches, in real-life scenarios remains that diverse weather conditions frequently manifest concurrently, such as the combination of rain and fog, which can render single-weather recovery methods ineffective. Therefore, multi-weather restoration techniques have been proposed. Li \textit{et al}. \cite{li2020all} employed neural architecture search to automatically optimize features from various weather encoders, aiming for uniform weather removal. Valanarasu \textit{et al}. \cite{valanarasu2022transweather} proposed an end-to-end transformer weather encoding-decoding architecture, enhancing the attention features within patches through intra transformer blocks. Chen \textit{et al}. \cite{chen2022learning} presented a unified framework for weather restoration by transfer knowledge from multiple teacher networks responsible for a specific weather recoveries to student networks. Wang \textit{et al}. \cite{wang2023smartassign} exploited the intrinsic correlation between rain and snow degradation to devise a knowledge allocation strategy, integrating transformer and Conv to architect the pipeline. Patil \textit{et al}. \cite{patil2023multi} developed a unified weather recovery model by employing a domain transformation technique that accommodates various weather conditions.

\subsection{De-weathering Datasets} \label{sec:2.2}
Most of the aforementioned approaches depend on manually synthesized datasets, such as Rain100L \cite{yang2017deep}, Rain100H \cite{yang2017deep}, RainCityscapes \cite{hu2019depth}, Rain12000 \cite{zhang2018density}, Rain14000 \cite{fu2017removing}, snow100k \cite{liu2018desnownet}, CSD \cite{chen2021all}, TransWeather \cite{valanarasu2022transweather}. These synthesis datasets primarily addresses the issue of limited availability of real-world data pairs by artificially superimposing rain, snow, fog, and other degradation effects onto clear images using sophisticated prior knowledge. Nevertheless, these manual patterns fail to accurately represent the degradation distributions in real-life, resulting in poor generalization performance. To address this issue, several recent studies introduced the multi-stage time dataset GT-RAIN \cite{ba2022not} that captured pairs of video frames, both with and without rain, during moments of minimal dynamic changes in the scene. The dataset is meticulously labeled to ensure feature consistency, with the sole exception being the rain effect. Zhang \textit{et al}. \cite{zhang2023weatherstream} improved this method by modeling the optical flow of scenes to automatically select time-multiplexed pairs, proposed a multi-weather dataset (WeatherStream) including rain, snow, and fog. These time-multiplexed real-world datasets are incredibly valuable to de-weathering research, yet there are instances where image details are misaligned, refer to Fig. \ref{fig:dataset_problems}.

\begin{figure}[t]
    \centering
    \includegraphics[width=0.9\linewidth]{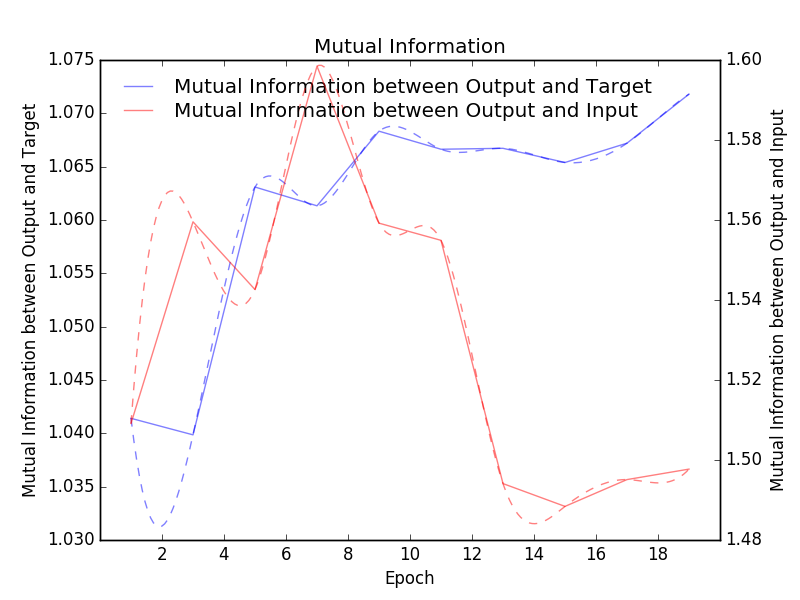}
    \vspace{-0.8em}
    \caption{The variation in mutual information during training phase. We compute the mutual information between the output and both the target and the input. The model is trained on GT-Rain-Snow \cite{ba2022not} dataset for a total of $20$ epochs, with mutual information being calculated every $2$ epochs.
    }
    \label{fig:mutual_info}
\end{figure}

\begin{figure*}[!ht]
    \centering
    \includegraphics[width=0.9\linewidth]{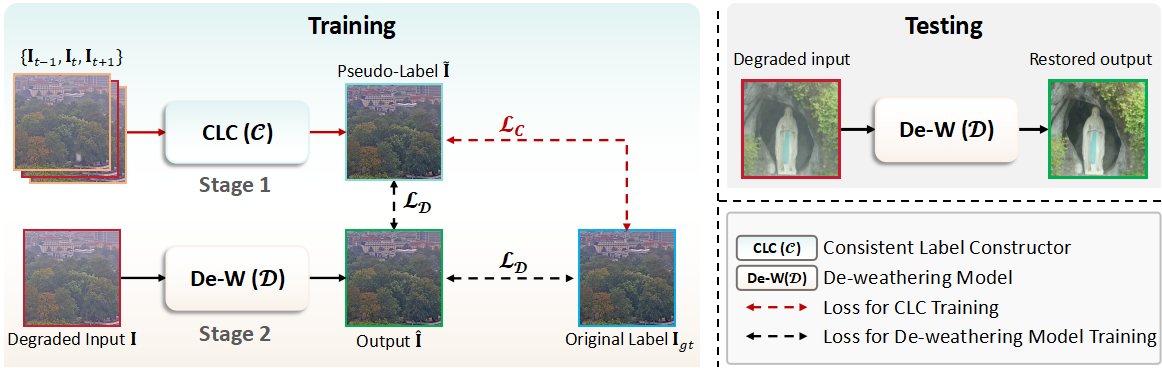}
    \caption{The pipeline of proposed method. A Consistent Label Constructor (CLC) is pre-trained to generate a pseudo-label $\mathbf{\tilde{I}}$ from multiple degraded frames. Then both the original label $\mathbf{I}_{gt}$ and the pseudo-label $\mathbf{\tilde{I}}$ are employed to regulate the output $\mathbf{\hat{I}}$ of the De-weathering Model (De-W). For testing, solely the De-W is utilized to restore the weather degraded input.
    }
  \label{fig:clc_pipe}
\end{figure*}
\section{Methodology} 
\label{sec:3}
\subsection{Motivation and Overall Pipeline} 
\label{sec:3.1}
Given an image $\mathbf{I}$ degraded by severe weather (\emph{e.g.}, rain, fog, and snow), the process of de-weathering can be represented as:
\begin{equation} 
\label{eq:3.1-1}
\mathbf{\hat{I}} = \mathcal{D} (\mathbf{I}; \Theta_\mathcal{D}),
\end{equation}
where $\Theta_\mathcal{D}$ denotes the parameters of the de-weathering model $\mathcal{D}$, which is typically attained by the following optimization objective in cooperation with the corresponding clean ground-truth (GT) image $\mathbf{I}_{gt}$,
\begin{equation}
    \Theta_\mathcal{D}^\ast = \arg\min_{\Theta_\mathcal{D}} \mathbb{E}_{\mathbf{I},\mathbf{G}} \left[ \mathcal{L}(\mathbf{\hat{I}}, \mathbf{I}_{gt}) \right],
    \label{eqn:de-wea}
\end{equation}
where $\mathcal{L}$ represents the supervised loss function.
However, in practice, the GT image is difficult or almost impossible to
acquire since that $\mathbf{I}_{gt}$ and $\mathbf{I}$ are not possible to capture simultaneously.
A straightforward solution is to utilize synthetic data for the training process; however, the domain gaps between the degradation model employed during training and that observed in actual images limits its efficacy.
While it is feasible to collect data from live streaming videos, it is inevitably influenced by the uncontrollable weather changes, resulting in inconsistencies in aspects like illumination, spatial position, and texture between $\mathbf{I}_{gt}$ and $\mathbf{I}$.
Despite the setting of strict filtering strategies and selection conditions in recent methods~\cite{ba2022not, zhang2023weatherstream}, it remains challenging to obtain completely ideal supervised images and there still exist some inconsistencies, as shown in Fig.~\ref{fig:dataset_problems}, which definitely brings adverse effects
on supervised learning of de-weathering model $\Theta_\mathcal{D}$.

Indeed, inconsistencies are also present in other computer vision tasks, such as super-resolution~\cite{zhang2022self,wang2023benchmark}, depth-estimation~\cite{li2023learning}, learnable ISP~\cite{zhang2021learning}, etc.~\cite{feng2023generating}.
To resolve inconsistent illumination, guided filter~\cite{he2012guided} is often adopted to correct output color to target, and then calculate the loss value~\cite{wei2020learning}.
For spatial misalignment, optical flow estimation~\cite{sun2018pwc} is extensively employed for the alignment of pairs, with certain studies even propose more robust loss functions, \emph{e.g.}, CoBi~\cite{kim2013cobi} and misalignment-tolerate $\ell_1$~\cite{xia2023image} loss.
Regrettably, these approaches exhibit decreased performance when applied to de-weathering tasks, attributable to the complexity and variability inherent in inconsistent weathering types.

In this research, we aim to explore a comprehensive approach for the effective training of de-weathering models amidst the diverse inconsistencies found in real-world de-weathering datasets. Drawing inspiration from the Information Bottleneck (IB) theory~\cite{tishby2015deep}, we introduce a fresh viewpoint to tackle this challenge. The IB theory posits that, during training, the mutual information between network features and the target rises steadily, whereas the mutual information between network features and the input initially rises and then falls, as illustrated in Fig. \ref{fig:mutual_info}. This indicates that the network progressively acquires target information, while it initially incorporates and later discards input information that is irrelevant to the target. This phenomenon provides us with the opportunity to generate pseudo-labels that closely align with the input image. Furthermore, in the context of de-weathering with imperfect supervision, addressing weather-irrelevant irregular disturbances in the target presents a greater challenge than mitigating weather-related degradation. Therefore, pseudo-labels are expected to eliminate most of the degradation without attempting to fit the imperfect parts of the original ground truth image.

Subsequently, we devise a Consistent Label Constructor (CLC) to produce pseudo-labels, incorporating multiple adjacent frames of the current degraded image to bolster the alignment between the pseudo-label and the degraded image, thereby enhancing the quality of the generated pseudo-label. It is important to acknowledge that relying solely on the pseudo-label as supervision for the de-weathering model is impractical, given the necessity of balancing consistency preservation with degradation removal. To address this, we integrate the pseudo-label with the original imperfect labels and retrain the de-weathering model using our proposed Information Allocation Strategy (IAS), which is elaborated upon in the subsequent section. The overview pipeline is presented in Fig.~\ref{fig:clc_pipe}.

\subsection{Consistent Label Constructor} \label{sec:3.2}
Given a de-weathering training set consisting of pairs of degraded image $\mathbf{I}$ and clean GT image $\mathbf{I}_{gt}$, when the supervision of the clean image is imperfect, we aim to assist learning of de-weathering model by generating pseudo-labels through our designed Consistent Label Constructor (CLC), \emph{i.e.},
\begin{equation}
\mathbf{\tilde{I}} = \mathcal{C} (\mathbf{I}; \Theta_\mathcal{C}).
\label{eqn:clc_1}
\end{equation}
The parameters $\Theta_\mathcal{C}$ of CLC model $\mathcal{C}$ can be optimized by the constraints from the original label $\mathbf{I}_{gt}$ as follows,
\begin{equation}
    \Theta_\mathcal{C}^\ast = \arg\min_{\Theta_\mathcal{C}} \mathbb{E}_{\mathbf{I},\mathbf{I}_{gt}} \left[\mathcal{L_\mathcal{C}}(\mathbf{\tilde{I}}, \mathbf{I}_{gt}) \right],
    \label{eqn:clc_p}
\end{equation}
where $\mathcal{L}_\mathcal{C}$ denotes the loss functions of CLC.
Fortunately, the existing real-world datasets are often collected from live streams, which enables the utilization of multiple adjacent frames to provide more information for pseudo-label generation.
The most intuitive method is to concatenate multiple frames as input, which is the approach we used in our conference version. Supposing that the current degraded image $\mathbf{I}$ is captured at time $t$, \emph{i.e.}, $\mathbf{I}_{t}$, we directly feed its neighboring $2n$ frames, \emph{i.e.}, $[\mathbf{I}_{t-n},\ldots,\mathbf{I}_{t-1},\mathbf{I}_{t},\mathbf{I}_{t+1},\ldots,\mathbf{I}_{t+n}]$, into the CLC model.
Meanwhile, it does not hinder the CLC from generating a pseudo-label $\mathbf{\tilde{I}_{t}}$ that aligns with $\mathbf{I}_t$, since the weather independent content remains consistent across frames captured within a brief interval.
Moreover, the utilization of complementary multi-frame information assists CLC in producing more faithful pseudo-labels, as rain streaks and snowflakes do not always remain in the same location across different frames.
As for the loss function in Eq. (\ref{eqn:clc_p}), we adopt $\ell_1$ as well as Multiscale Structural Similarity (MS-SSIM \cite{wang2003multiscale}) distance between the generated pseudo-label $\mathbf{\tilde{I}}$ and the original label $\mathbf{I}_{gt}$, which can be written as:
\begin{equation}
\mathcal{L_\mathcal{C}}(\mathbf{\tilde{I}}, \mathbf{I}_{gt})=\left\|\mathbf{\tilde{I}}-\mathbf{I}_{gt}\right\|_{1}+ \left(1-\mathcal{L}_{SSIM}\left(\mathbf{\tilde{I}}, \mathbf{I}_{gt} \right) \right)
\label{eqn:clc_c}
\end{equation}

In the conference version, we utilize RainRobust \cite{ba2022not} and Restormer \cite{zamir2022restormer} as the backbone networks for CLC. Despite the generated pseudo-labels maintaining consistency with the input in terms of content, the removal of weather degradation is suboptimal. 
The approach of directly concatenating multiple frames as input is overly simplistic, preventing the model from fully leveraging temporal information to generate high-quality pseudo-labels.
Therefore, in this manuscript, we design a novel module, \emph{i.e.}, Cross-frame Similarity Aggregation (CSA), to integrate multi-frame information and incorporate it into the CLC model to enhance the generation capability of high-quality pseudo-labels.

\subsection{Cross-frame Similarity Aggregation} 
\label{sec:3.3}
\begin{figure*}[ht]
    \centering
    \includegraphics[width=0.9\linewidth]{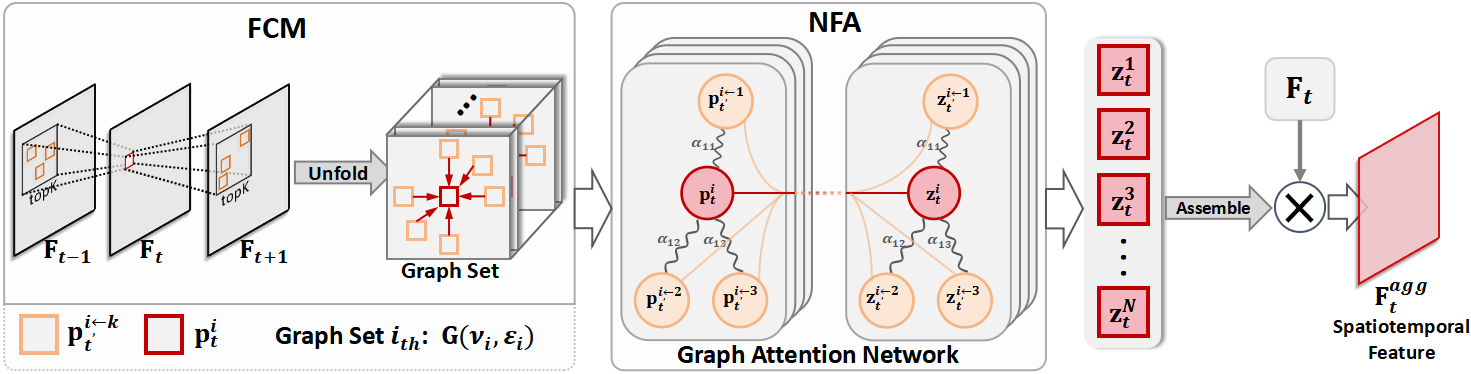}
    \caption{Details of the Cross-frame Similarity Aggregation (CSA) module, including Flexible Cross-frame Matching (FCM) and Non-local Feature Aggregation (NFA).}
    \label{fig:csa_pros}
\end{figure*}
To fully leverage the complementary information between multiple adjacent frames, we design the Cross-frame Similarity Aggregation (CSA) module, which primarily aggregates highly matched self-similar feature patches from sequential inputs using a Graph Attention Network (GAT)\cite{velivckovic2017graph}.

According to~\cite{zontak2011internal}, natural images contain numerous recurring redundant data patches, which provide powerful image-specific priors for image reconstruction problems.
The time-multiplexed real-world datasets~\cite{ba2022not,zhang2023weatherstream} are sourced from fixed cameras and undergo rough alignment operation. 
As a result, the sequences of images captured from the same scene at different timestamps also exhibit highly similar patch cross frames as illustrated in Fig. \ref{fig:adj_patch}. 
Therefore, we aim to explore the intrinsic prior knowledge of data by collecting the cross-frame self-similarity patches to aggregate non-local features, thereby enhancing the representation ability of the model. 
To achieve this, we refer to previous works~\cite{wang2018non,zhang2019residual,zhou2020cross}, which have been extensively validated in the field of image restoration and the non-local operation can be generally represented as:
\begin{equation}
{\mathbf{y}}_{i}=\frac{1}{Z (\mathbf{x})}\sum_{\forall j} f(\mathbf{x}_{i},\mathbf{x}_{j}) g(\mathbf{x}_{j})
\label{eq:4.1_1}
\end{equation}
where $i$ is the index of an output feature position and $j$ is the index that enumerates all possible positions. $\mathbf{x}$ is the input image feature and $\mathbf{y}$ is the output signal of the same size as $\mathbf{x}$. $f(\cdot,\cdot)$ is the similarity measurement function, and $g(\cdot)$ computes a representation of the input at the position $j$. $Z(\cdot)$ denotes normalization factor. 

%
Drawing inspiration from the non-local operation in Eq. (\ref{eq:4.1_1}), we design the Cross-frame Similarity Aggregation (CSA) in feature space and the module is presented in Fig. \ref{fig:csa_pros}, which includes Flexible Cross-frame Matching (FCM) and Non-local Feature Aggregation (NFA).
The process of FCM aims to locate the self-similar patches in adjacent frames and NFA is employed to aggregate the non-local features.
In particular, we represent similar patches across frames using a graph structure~\cite{scarselli2008graph} within the FCM and employ Graph Attention Networks (GAT) \cite{velivckovic2017graph} for feature fusion learning in the NFA.

It is worth noting that the reason for not directly adopting the conventional non-local network for feature aggregation here lies in its reliance on predefined algorithms to compute correlation weights that guide the aggregation process. In contrast, the GAT dynamically updates these correlation weights based on the latent relationships between nodes. This approach facilitates a more accurate representation of the complex degradation patterns present in real-world de-weathering tasks. The subsequent Sec. \ref{sec:5.6.4} provides an experimental analysis across various feature aggregation methods.

\subsubsection{Flexible Cross-frame Matching (FCM)} 
\label{sec:3.3.2}
Given the current degraded image $\mathbf{I}_{t}$, we consider its neighboring frames $\{ \mathbf{I}_{t-1}, \mathbf{I}_{t+1} \}$ and extract their feature representations denoted as $\{\mathbf{F}_{t-1}, \mathbf{F}_{t}, \mathbf{F}_{t+1} \}$.
A $s\times s$ sliding window is applied to extract a number of $N$ query patches from $\mathbf{F}_{t}$ and locate self-similar patches in adjacent frames that can be matched with them.
The specific procedure is illustrated in Fig. \ref{fig:csa_pros}. 
All query patches are unfolded from the current frame feature $\mathbf{F}_t$, resulting in $\{\mathbf{p}^{i}_{t} \in \mathbb{R}^{C \times s^2}\}_{i=1}^{N}$, where $C$ represents the channel number of $\mathbf{F}_t$.
Then, the flexible matching is performed within an extended region according to the position of query patch $\mathbf{p}^{i}_{t}$ among all adjacent frames $\mathbf{F}_{t'}$ with $\ {t'\in [t-n,t+n]\setminus\left \{ t \right \}}$. 
The extended matching region is determined by a pre-set Region Padding Size ($\text{P}$), resulting in a total expanded area of $(s+2\text{P})^{2}$. 
Within the search regions, we can extract a total of $M$ patches from features of adjacent frames, \emph{i.e.}, $\{\mathbf{p}^{j}_{t'} \in \mathbb{R}^{C \times s^2}\}_{j=1}^{M}$.
Considering the unacceptable computational cost of aggregating features for all patches $\mathbf{p}^{j}_{t'}$ within the extended region of adjacent frames with the query patch $\mathbf{p}^{i}_{t}$, we propose an optimization strategy: to calculate the cosine similarity between the query patch $\mathbf{p}^{i}_{t}$ and each of the M patches $\mathbf{p}^{j}_{t'}$.
\begin{equation}
{f}_{ cos\_sim }(\mathbf{p}^{i}_{t}, \mathbf{p}^{j}_{t'}) = \frac{\mathbf{p}^{i}_{t}\cdot \mathbf{p}^{j}_{t'}}{ \| \mathbf{p}^{i}_{t}  \| \times  \| \mathbf{p}^{j}_{t'}  \| }.
\label{eq:4.2_1}
\end{equation}
According to the calculated cosine similarity, we then select the top-$K$ similar patches for the query patch $\mathbf{p}^{i}_{t}$ from adjacent frames, denoted as 
$\{\mathbf{p}^{i\leftarrow k}_{t'} \in \mathbb{R}^{C \times s^2}\}_{k=1}^{K}$ with $k \in [1, M]$, where $\leftarrow$ indicates the relationship between those non-local features with the query patch.
Then, to enable the adaptive aggregation of non-local features, we represent the involved patch features in the form of graphs and treat each patch as a node in the graph structure. 
Adjacent frame patches are directed towards the query patch ($\mathbf{p}^{i}_{t}\leftarrow \mathbf{p}^{i\leftarrow k}_{t'}$), giving rise to the edge matrix $\mathcal{E}$ and ultimately yielding the directed graph set $\mathcal{G} = \bigcup_{i=1}^{N} \mathbf{G}_i, \text{where } \mathbf{G}_i = (\mathcal{V}_i, \mathcal{E}_i)$. 
Afterwards, the constructed graph set with dimensions denoted as (B, K, N, C$\times s^{2}$) are subsequently fed into the GAT to facilitate feature aggregation, where B refer to the batch size, respectively. GAT will process each batch sequentially, treating the entire graph set $\mathcal{G}$ as new batch of data with a size of K for efficient processing.

\subsubsection{Non-local Feature Aggregation (NFA)} 
\label{sec:3.3.3}
To aggregate the node features, the attention coefficient is required to quantify the associative weights between query node and its adjacent nodes.
For each sub-graph $\mathbf{G}_i$, we calculate the graph attention coefficient of matched nodes $k$ to query node $i$ by the following formula according to GAT \cite{velivckovic2017graph}:
\begin{equation}
\alpha_{ik} = \frac{ \exp \left( \rho \left({\mathbf{a}}^{T} \left[ \mathbf{W} {\mathbf{p}}_{t}^{ i} \| \mathbf{W} {\mathbf{p}}_{t'}^{i\leftarrow k)} \right] \right) \right)}
{\sum_{k\in \mathcal{N}_{i}} \exp \left( \rho \left({\mathbf{a}}^{T}[\mathbf{W}{\mathbf{p}}^{i}_{t} \| \mathbf{W} {\mathbf{p}}^{i\leftarrow k}_{t'} \right) \right) }
\label{eq:4.2_2}
\end{equation}
where $\mathbf{W} $ and $\mathbf{a}$ denote the learnable weight matrix and vector, respectively. $\rho(\cdot)$ refers to the LeakyReLU operation. $\mathcal{N}_{i}$ denotes the set of all matched adjacent frame nodes of node $i$. 
The correlation between the query node and the adjacent frame nodes within the feature space can be accurately estimated, which in turn provides weighted coefficients for the subsequent feature aggregation process. 
Based on the attention coefficients $a_{ij}$ obtained from Eq. (\ref{eq:4.2_2}), the cross-frame non-local feature aggregation process can be expressed as:
\begin{equation}
{\mathbf{z}}^{i}_{t}=\phi (\sum_{k\in \mathcal{N}_{i}} \alpha_{ik} \mathbf{W}{\mathbf{p}}^{i\leftarrow k}_{t'})
\label{eq:4.2_3}
\end{equation}
where $\phi$ represents the activation function and $\mathbf{z}^{i}_{t}$ is the obtained updated node feature by aggregating the features of all neighboring nodes. 
After achieving these query nodes $\mathbf{z}^{i}_{t},i=\left \{ 1,\dots ,N \right \}$ through GAT, We then reassemble them in the order of unfolding, followed by element-wise multiplication with the input feature map $\mathbf{F}_{t}$, which has the same dimensions, yielding a spatiotemporal aggregated feature $\mathbf{F}^{agg}_{t}$.

Since the degradation of images depicting the same scene due to weather conditions such as rain and snow may exhibit nuances over time, harnessing cross-frame self-similarity and FCM allows us to identify highly correlated and minimally degraded non-local features from neighboring frames. This approach facilitates the restoration of original regions obscured by rain and snow using specific prior information.

\begin{figure*}[!h]
    \centering
    \includegraphics[width=0.9\linewidth]{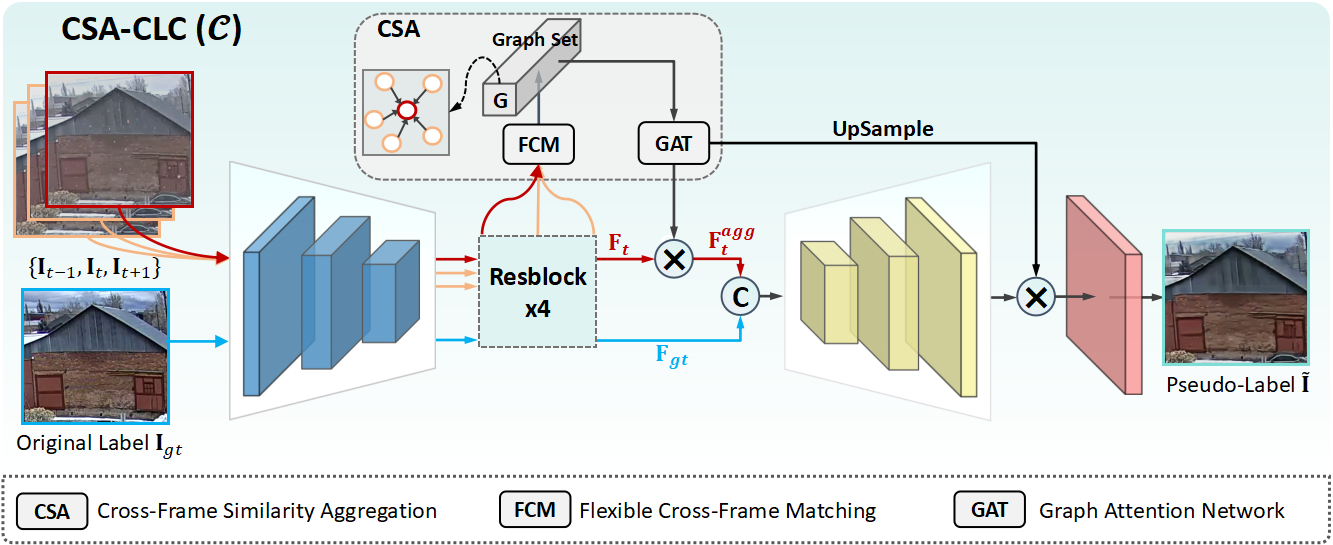}
    \caption{This figure illustrates the current frame along with its preceding and succeeding frames collectively serving as the degraded input for the CSA-CLC, with the Input Frame Number $\text{N}=3$. CSA-CLC participates in training only, where it processes multi-frame $\left \{\mathbf{I}_{t\pm1}, \mathbf{I}_{t} \right \}$ and original labels $\mathbf{I}_{gt}$. The encoder downsamples all inputs into feature space using shared parameters. Subsequently, cross-frame self-similar features are aggregated through the CSA. Specifically, a feature correlation graph set is constructed via FCM, which is then fed into GAT for feature aggregation, resulting in the spatiotemporal aggregated feature $\mathbf{F}^{agg}_{t}$. Following steps involve aligning the feature of degraded input with label $\mathbf{F}_{gt}$ and then upsampling to generate aligned labels.}
    \label{fig:csa-clc_pipe}
\end{figure*}

\subsection{The Improved Consistent Label Constructor} 
\label{sec:3.4}
In this section, we integrate the proposed Cross-frame Similarity Aggregation (CSA) module into the original Consistent Label Constructor (CLC) network, yielding an improved CSA-CLC model that generates higher-quality pseudo-labels.
It should be noted that the CSA-CLC network is exclusively involved in training and is not utilized during the testing phase. 
It is designed to generate pseudo-labels that are consistent with the degraded image in non-weather content while removing degradation caused by adverse weather conditions.
This allows us to use the non-ideal ground truth (GT) images as inputs to the network, providing a clean signal reference that enhances the ability of CSA-CLC to remove degradation and produce high-quality pseudo-labels.
The pipeline of CSA-CLC model is illustrated in Fig. \ref{fig:csa-clc_pipe}.
Compared with the original CLC in our conference version, the improved CSA-CLC thoroughly exploits the spatio-temporal information within the multi-frame inputs, and significantly enhances the quality and fidelity of the generated pseudo-labels, enabling the de-weathering model to achieve a satisfactory performance enhancement.
\noindent{\textbf{Network Architecture.}} 
Our CSA-CLC employs the standard encoder-decoder architecture. 
The inputs include the degraded image $\mathbf{I_t}$, its preceding and subsequent $n$ adjacent frames $\left \{\mathbf{I}_{t-n}, \ldots, \mathbf{I}_{t}, \ldots, \mathbf{I}_{t+n}\right \}$, as well as the original non-ideal label $\mathbf{I}_{gt}$. These inputs are mapped into the feature space through shared parameters, yielding the feature representations, \emph{i.e.}, $\left \{\mathbf{F}_{t-n}, \ldots, \mathbf{F}_{t}, \ldots, \mathbf{F}_{t+n}, \mathbf{F}_{gt} \right \} $. 
Subsequently, we utilize the Flexible Cross-frame Matching (FCM) module to construct collections of cross-frame feature patches from $\left \{\mathbf{F}_{t-n}, \ldots, \mathbf{F}_{t}, \ldots, \mathbf{F}_{t+n} \right \} $, which are represented in the form of directed graphs. 
Next, we deploy a Graph Attention Network (GAT) consisting of two layers to perform non-local feature aggregation on each encapsulated directed graph, thereby obtaining the ensemble feature $\hat{\mathbf{F}}_{t}$, This feature is then multiplied element-wise with $\mathbf{F_t}$ to obtain $\mathbf{F}^{agg}_{t}$.
Finally, the resulting feature is further concatenated with the GT image feature $\mathbf{F}_{gt}$ along the feature dimension and fed into the subsequent decoder network to generate the pseudo-label image.
The encoder-decoder architecture and the Resblocks are instantiated with convolutional layers, each accompanied by batch normalization (BN) and ReLU operations, while the final layer employs the Tanh activation function. 
\noindent{\textbf{Learning Objective.}}
In accordance with the optimization objective of CLC, \emph{i.e.}, $\mathcal{L}_{\mathcal{C}}$ in Eq. (\ref{eqn:clc_c}), we incorporate a Rain-robust loss $\mathcal{L}_{R}$~\cite{ba2022not} in this version and obtain the following loss function $\mathcal{L}_\mathcal{C^{*}}$ for CSA-CLC:

\begin{equation}
\mathcal{L}_\mathcal{C^{*}}(\mathbf{\tilde{I}}, \mathbf{I}_{gt})= \mathcal{L}_{\mathcal{C}}(\mathbf{\tilde{I}}, \mathbf{I}_{gt}) + \mathcal{L}_{R}(\mathbf{\tilde{I}}, \mathbf{I}_{gt})
\label{eqn:csaclc_c}
\end{equation}
\begin{figure}[ht]
    \centering
    \includegraphics[width=1\linewidth]{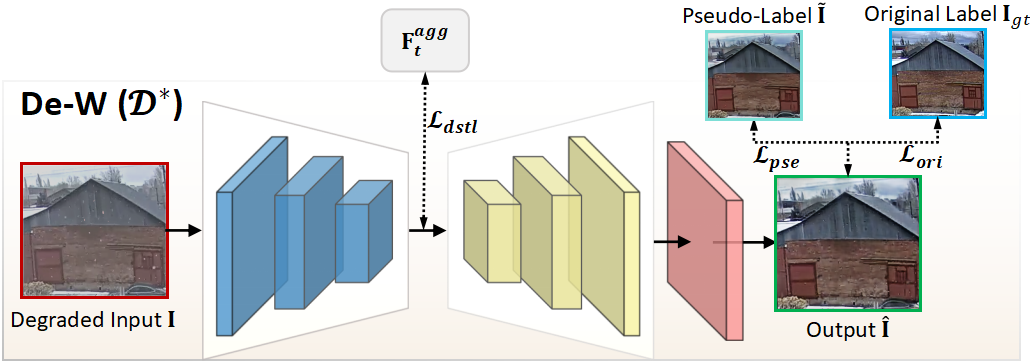}
    \vspace{-4mm}
    \caption{The overall IAS of CSA-CLC pipeline. De-W achieves de-weathering learning through the joint supervision of pseudo-label, original label and spatiotemporal aggregation feature.}
    \label{fig:dew_supv}
\end{figure}
\subsection{Information Allocation Strategy} 
\label{sec:3.5}
Although the pseudo-labels are more consistent with the inputs than the original labels, there may still be issues with insufficient degradation removal, inevitably impacting the subsequent learning of the de-weathering model.
Therefore, instead of relying solely on pseudo-labels as supervisory signals, we also include the original labels in a co-supervised manner.
To achieve this, we propose an Information Allocation Strategy (IAS) to balance and combine the advantages of the pseudo-label $\mathbf{\tilde{I}}$ and its corresponding original label $\mathbf{I}_{gt}$ in our conference version.
To maintain consistency with input, we directly apply the image-level constraint, \emph{i.e.}, $L_1$ loss, between the output of de-weathering model $\mathbf{\hat{I}}$ and the pseudo-label $\mathbf{\tilde{I}}$.
In order to extract the pristine information from the initial labels $\mathbf{I}_{gt}$ while simultaneously reducing the effects of their inconsistencies to the fullest extent, relying on pixel-level loss functions is inadequate. As a result, we employ constraints at both the feature-level and distribution-level, owing to their superior robustness against spatial misalignments and color variations that may occur between corresponding image pairs.

In our previous work (CLC pipeline), we impose constraints using the Rain-robust loss~\cite{ba2022not} and the Sliced Wasserstein (SW) loss~\cite{zhang2022self}.
The Rain-robust loss, a variant of contrastive loss, is computed based on the intermediate features extracted from de-weathering model.
SW loss is a probability distribution distance between VGG~\cite{simonyan2014very} features.
Using the IAS, we then formulate the loss function for the learning of de-weathering model in CLC pipeline as:
\begin{equation}
\mathcal{L_\mathcal{D}}(\mathbf{\hat{I}}, \mathbf{\tilde{I}}, \mathbf{I}_{gt}) \!=\!
\mathcal{L}_{1}(\mathbf{\hat{I}},\mathbf{\tilde{I}}) \!+\! \lambda_{1} (\mathcal{L}_{R}(\mathbf{\hat{I}}, \mathbf{I}_{gt}) \!+\! \lambda_{2}\mathcal{L}_{SW}(\mathbf{\hat{I}}, \mathbf{I}_{gt}) )
\label{eq:loss_las}
\end{equation}
where $\lambda_{1}$ and $\lambda_{2}$ denote the trade-off of the supervision from the original label and SW, $\mathcal{L}_{R}$ and $\mathcal{L}_{SW}$ denote the terms of Rain-robust and SW loss, respectively.
In Eq. (\ref{eq:loss_las}), we empirically set $\lambda_{1}=0.1$ and $\lambda_{2}=0.08$.
The details and formulas of $\mathcal{L}_{Robust}$ and $\mathcal{L}_{SW}$ are provided in the supplementary material.
In this manuscript, we enhance the CLC pipeline and propose the CSA-CLC, which not only promotes the generation of higher-quality pseudo-labels, but also strengthens feature representation through the multi-frame fusion strategy. 
To fully leverage CSA-CLC, we also implement several adjustments to the previous IAS.
When using pseudo-labels, we incorporate not only the image-level constrain, \emph{i.e.}, $\mathcal{L}_1$, but also introduce the multi-scale SSIM loss $\mathcal{L}_{SSIM}$~\cite{wang2003multiscale} as follows,
\begin{equation}
\mathcal{L}_{pse}(\mathbf{\hat{I}}, \mathbf{\tilde{I}})=\mathcal{L}_{1}+(1-\mathcal{L}_{SSIM}(\mathbf{\hat{I}}, \mathbf{\tilde{I}}))
\end{equation}
This enables IAS to leverage the clear structural details in $\mathbf{\tilde{I}}$, thereby providing more efficient guidance for the de-weathering output $\mathbf{\hat{I}}$.
As for the original labels, we observe that replacing the SW loss with the MS-SSIM loss leads to the learning of de-weathering models with comparable performance, while also simplifying the training process.
Thus, we simplify the supervision of original label to:
\begin{equation}
\mathcal{L}_{ori}(\mathbf{\hat{I}}, \mathbf{I}_{gt})=\mathcal{L}_{R}(\mathbf{\hat{I}}, \mathbf{I}_{gt})+(1-\mathcal{L}_{SSIM}(\mathbf{\hat{I}}, \mathbf{I}_{gt}))
\end{equation}
To take advantage of the enhanced feature representation of CSA-CLC, we also conduct feature distillation in the latent space. We enforce a feature-level constraint utilizing the smooth $\mathcal{L}_{1}$ loss as
\begin{equation}
\mathcal{L}_{dstl} = \mathcal{L}^{smooth}_1(\mathbf{G}_{t},\text{stopgrad}(\mathbf{F}^{agg}_{t}))
\label{eq:4.3_3}
\end{equation}
where $\mathbf{F}_{t}$ denotes the feature map of input, $\mathbf{F}^{agg}_{t}$ represents the aggregated feature map after CSA process, and $\mathbf{G}_{t}$ is the intermediate feature map of de-weathering model (De-W).
As shown in \ref{fig:dew_supv}, the learning objective of de-weathering model equipped with the proposed CSA-CLC can be presented as:
\begin{equation}
\mathcal{L_\mathcal{D^{*}}}(\mathbf{\hat{I}}, \mathbf{\tilde{I}}, \mathbf{I}_{gt})=\mathcal{L}_{pse}(\mathbf{\hat{I}}, \mathbf{\tilde{I}})+\lambda_{o}\mathcal{L}_{ori}(\mathbf{\hat{I}}, \mathbf{I}_{gt})+\lambda_{d}\mathcal{L}_{dstl}
\label{eqn:csaclc_d}
\end{equation}
and we empirically set $\lambda_{o}=0.25, \lambda_{d}=0.01$.

\section{Experiments} 
\label{sec:5}
\subsection{Datasets} 
\label{sec:5.1}
GT-RAIN~\cite{ba2022not} pioneered the use of time multiplexed pairs to construct a high-quality dataset of real rainy and clean image pairs. However, it is limited to a single weather type (rain) and suffers from scalability constraints due to its reliance on manual annotation.
Building on the protocol of GT-RAIN, WeatherStream~\cite{zhang2023weatherstream} extends the framework to multiple weather conditions, such as rain, snow, and fog, through automated data collection. 
Additionally, GT-RAIN-SNOW\href{https://drive.google.com/drive/folders/1_-EPWeTjrpg7EICz93uVaOkS_PIHz5uW}{$^{1}$} integrates the GT-RAIN and GT-SNOW datasets to address both rainy and snowy conditions, where GT-SNOW is an auxiliary dataset contributed by \cite{zhang2023weatherstream}.
In this work, we leverage two real-world time-multiplexed datasets: GT-RAIN-SNOW and WeatherStream, both of which exhibit inherent label misalignment issues due to temporal and environmental inconsistencies.
Specifically, GT-RAIN-SNOW encompasses 129 distinct scenes, whereas WeatherStream features a broader array of 424 scenes. Each scene within these datasets comprises roughly 300 aligned degraded images, accompanied by a single clean reference image.
We conduct experiments using these two training datasets independently and evaluate performance on the testing dataset from WeatherStream, which includes 45 scenes for sub-tasks, including deraining, dehazing, and desnowing.

\begin{table*}[!t]
\footnotesize
\centering
\caption{Quantitative comparison of de-weathering results from CLC pipeline and CSA-CLC pipeline with other state-of-the-art methods on GT-RAIN-SNOW and WeatherStream datasets. \label{tab:5.3_1}}
\resizebox{0.95\textwidth}{!}{
\begin{tabular}{clcccccccc}
\hline
\multirow{2}{*}{Dataset}                                  & \multirow{2}{*}{Method}              & \multicolumn{2}{c}{Rain}         & \multicolumn{2}{c}{Fog}          & \multicolumn{2}{c}{Snow}         & \multicolumn{2}{c}{Overall}      \\ \cline{3-10} 
                                                          &                                      & PSNR$\uparrow$ & SSIM$\uparrow$  & PSNR$\uparrow$ & SSIM$\uparrow$  & PSNR$\uparrow$ & SSIM$\uparrow$  & PSNR$\uparrow$ & SSIM$\uparrow$  \\ \hline
\multirow{7}{*}{GT-RAIN-SNOW \cite{ba2022not}}                   & TransWea \cite{valanarasu2022transweather}     & 21.84          & 0.7635          & 21.44          & 0.7624          & 21.51          & 0.7940          & 21.63          & 0.7731          \\
                                                          & Uformer \cite{wang2022uformer}      & 22.37          & 0.7757          & 19.03          & 0.7459          & 21.06          & 0.7872          & 20.84          & 0.7672          \\
                                                          & DSRformer \cite{chen2023learning}   & 22.97          & 0.7928          & 21.14          & 0.7794          & 22.15          & 0.8061          & 22.12          & 0.7929          \\
                                                          & Histoformer \cite{sun2025restoring} & 22.90          & 0.7892          & 20.98          & 0.7763          & 22.07          & 0.8053          & 21.98          & 0.7912          \\ \cline{2-10} 
                                                          & Ours(CLC)                            & \textbf{23.24} & \textbf{0.7980} & \textbf{21.52} & \textbf{0.7860} & \textbf{22.31} & \textbf{0.8107} & \textbf{22.40} & \textbf{0.7977} \\
                                                          \rowcolor{mycolor!25} \multirow{-6}{*}{\cellcolor{white}{}} & Ours(CSA-CLC)                     & \textbf{23.47} & \textbf{0.8007} & \textbf{21.59} & \textbf{0.7874} & \textbf{22.38} & \textbf{0.8183} & \textbf{22.51} & \textbf{0.8004} \\ \hline
\multirow{7}{*}{WeatherStream \cite{zhang2023weatherstream}} & TransWea \cite{valanarasu2022transweather}     & 22.43          & 0.7873          & 22.57          & 0.7740          & 21.83          & 0.7921          & 22.27          & 0.7846          \\
                                                          & Uformer \cite{wang2022uformer}      & 22.41          & 0.7908          & 20.03          & 0.7711          & 21.15          & 0.8012          & 21.23          & 0.7879          \\
                                                          & DSRformer \cite{chen2023learning}   & 23.52          & 0.7964          & 22.55          & \textbf{0.7973} & 22.29          & 0.8148          & 22.83          & 0.8022          \\
                                                          & Histoformer \cite{sun2025restoring} & 23.49          & 0.7967          & 22.64          & 0.7956          & 22.27          & 0.8152          & 22.86          & 0.8019          \\ \cline{2-10} 
                                                          & Ours(CLC)                            & \textbf{23.68} & \textbf{0.7987} & \textbf{22.71} & 0.7971          & \textbf{22.54} & \textbf{0.8163} & \textbf{23.02} & \textbf{0.8033} \\
                                                          \rowcolor{mycolor!25} \multirow{-6}{*}{\cellcolor{white}{}} & Ours(CSA-CLC)      & \textbf{23.93} & \textbf{0.8058} & \textbf{23.22} & \textbf{0.8061} & \textbf{22.78} & \textbf{0.8237} & \textbf{23.38} & \textbf{0.8106} \\ \hline
\end{tabular}
}
\end{table*}

\begin{figure*}[ht]
  \centering
  \captionsetup[subfloat]{labelformat=empty}
  {\includegraphics[width=0.95\textwidth]{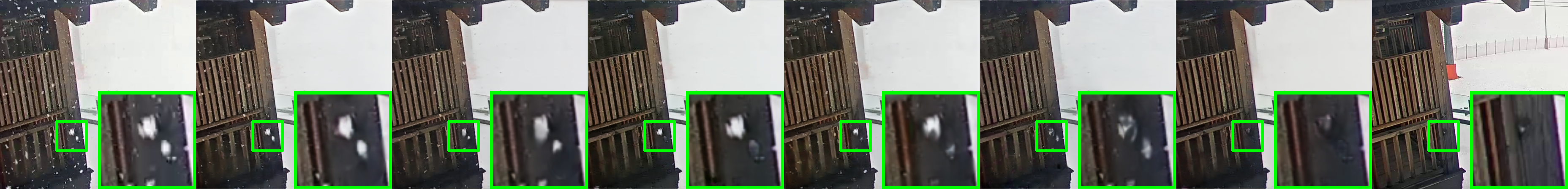}}
  {\includegraphics[width=0.95\textwidth]{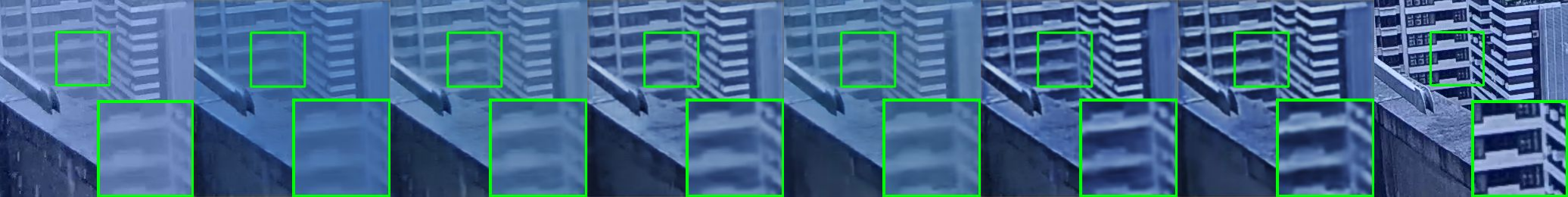}}\vspace{-1em}
  \subfloat[\raggedright\footnotesize\qquad\ \ Input\qquad\qquad Uformer\cite{wang2022uformer}\qquad TransWea\cite{valanarasu2022transweather}\quad\  DSRformer \cite{chen2023learning}\ \ \ Histoformer\cite{sun2025restoring}\qquad Ours(CLC)\quad\ Ours(CSA-CLC)\qquad\ \ \ Clean]
  {\includegraphics[width=0.95\textwidth]{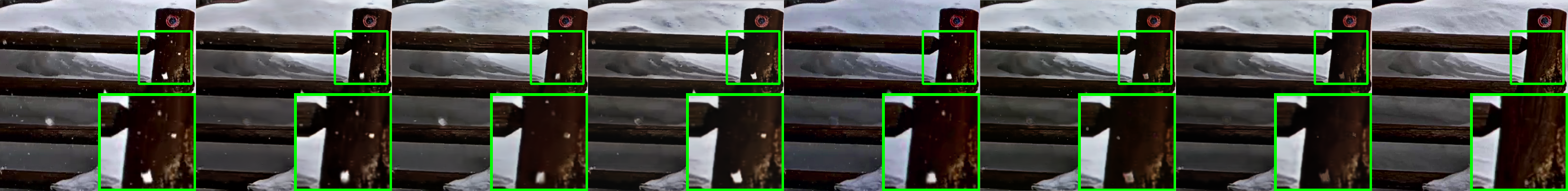}}
  \caption{Visual comparisons of De-weathering results from CLC pipeline and CSA-CLC pipeline with other state-of-the-art methods.}
  \label{fig:qlt_cpr}
\end{figure*}

\subsection{Implementation Details} 
\label{sec:5.2}
During training, we follow~\cite{ba2022not} to incorporate various augmentation techniques such as random rotation, padding, cropping, Rain-Mix \cite{guo2021efficientderain}, and Snow-Mix \cite{zhang2023weatherstream}. For optimization, we utilize the Adam optimizer with parameters $\beta_1 = 0.9$ and $\beta_2 = 0.999$. A warm-up strategy is implemented to progressively elevate the learning rate from $5\times10^{-5}$ to $2\times10^{-4}$ over the initial four epochs, succeeded by a cosine annealing decay to $10^{-6}$ throughout the remaining epochs.

Input images are randomly cropped to $256\times 256$, with $\text{batchsize}=8$. 
In addition, we set the details of the pipeline with CSA-CLC as follows: the input frame number is 3, the Region Padding Size $\text{P}=3$ and the number of selecting top-$K$ similarity patches for aggregation is 3. All experiments are implemented using PyTorch on RTX A6000 GPU.

\subsection{Comparison with State-of-the-Arts} 
\label{sec:5.3}
We select preceding state-of-the-art weather restoration methods, \emph{e.g.}, Uformer \cite{wang2022uformer}, TransWea \cite{valanarasu2022transweather}, DSRformer \cite{chen2023learning} and Histoformer \cite{sun2025restoring},  with a particular emphasis on those tailored for multi-weather image restoration. 
We calculate the averaged PSNR and SSIM metrics across multiple experiments, in conjunction with visual results, to conduct quantitative and qualitative analysis.

\subsubsection{Quantitative Comparisons}
For real-world de-weathering datasets with misalignment issues, our experimental results demonstrate that the proposed joint supervision architecture consistently outperforms existing methods in addressing weather-induced degradations, such as rain, snow, and fog, in most cases, as shown in Table \ref{tab:5.3_1}. 
When trained on WeatherStream, the CLC-based approach achieves slightly lower yet competitive SSIM result compared to DSRformer for dehazing task.
These results collectively validate that the alignment training mode significantly enhances the network's learning capability on real-world datasets.

Notably, by leveraging the improved CSA-CLC pipeline, our deweathering model achieves superior performance compared to the CLC-based approach. This improvement can be attributed to two key factors: (i) higher-quality pseudo-labels generated by CSA-CLC, (ii) the distillation of rich cross-frame aggregated features into de-weathering model.
By integrating multi-dimensional auxiliary information at both feature- and label-level, the CSA-CLC pipeline further boosts the performance of deweathring model, demonstrating its effectiveness in handling complex real-world de-weathering tasks.

\subsubsection{Qualitative Comparisons}
A qualitative comparison with the state-of-the art methods is provided in Fig. \ref{fig:qlt_cpr}. 
As illustrated in the first and third rows of results, existing methods often leave behind noticeable snowflake artifacts, whereas our approach demonstrates a more thorough elimination of these artifacts, particularly when using the CSA-CLC pipeline. 
For raindrop and fog degradation (second row), the images processed by our method exhibit significantly clearer outputs with minimal residual raindrops. Furthermore, by leveraging precise aggregation of cross-frame self-similar features, the CSA-CLC pipeline achieves superior preservation of texture and structural details, further enhancing the overall visual quality of the restored images.

\subsection{Evaluation on Different Backbones} 
\label{sec:5.4}
\begin{table*}[!t]
\centering
\caption{Quantitative comparisons of De-weathering results from our proposed pipeline with various backbone networks on the GT-RAIN-SNOW and WeatherStream datasets. \label{tab:5.4_1}}
\begin{tabular}{clllllllll}
\hline
{ }                                & { }                             & \multicolumn{2}{c}{{ Rain}}                                & \multicolumn{2}{c}{{ Fog}}                                 & \multicolumn{2}{c}{{ Snow}}                                & \multicolumn{2}{c}{{ Overall}}                             \\ \cline{3-10} 
\multirow{-2}{*}{{ Dataset}}       & \multirow{-2}{*}{{ Mothod}}     & { PSNR$\uparrow$} & { SSIM$\uparrow$}  & { PSNR$\uparrow$} & { SSIM$\uparrow$}  & { PSNR$\uparrow$} & { SSIM$\uparrow$}  & { PSNR$\uparrow$} & { SSIM$\uparrow$}  \\ \hline
{ }                                & { Restormer \cite{zamir2022restormer}} & { 22.63}          & { 0.7940}          & { 20.14}          & { 0.7700}          & { 21.62}          & { 0.8123}          & { 21.51}          & { 0.7913}          \\
{ }                                & { Ours(CLC)-Restormer}          & { \textbf{23.10}} & { \textbf{0.7971}} & { \textbf{21.46}} & { \textbf{0.7860}} & { \textbf{21.78}} & { \textbf{0.8178}} & { \textbf{22.17}} & { \textbf{0.7995}} \\
\rowcolor{mycolor!25} \multirow{-6}{*}{\cellcolor{white}{}} & { Ours(CSA-CLC)-Restormer}      & { \textbf{23.34}} & { \textbf{0.7992}} & { \textbf{21.63}} & { \textbf{0.7907}} & { \textbf{21.80}} & { \textbf{0.8176}} & { \textbf{22.28}} & { \textbf{0.8013}} \\ \cline{2-10} 
{ }                                & { RainRobust \cite{ba2022not}}         & { 22.83}          & { 0.7887}          & { 20.95}          & { 0.7691}          & { 22.17}          & { 0.8058}          & { 22.01}          & { 0.7871}          \\
{ }                                & { Ours(CLC)-RainRobust}         & { \textbf{23.24}} & { \textbf{0.7980}} & { \textbf{21.52}} & { \textbf{0.7860}} & { \textbf{22.31}} & { \textbf{0.8107}} & { \textbf{22.40}} & { \textbf{0.7977}} \\
\rowcolor{mycolor!25} \multirow{-6}{*}{\cellcolor{white}{GT-RAIN-SNOW \cite{ba2022not}}} & { Ours(CSA-CLC)-RainRobust}     & { \textbf{23.47}} & { \textbf{0.8007}} & { \textbf{21.59}} & { \textbf{0.7874}} & { \textbf{22.38}} & { \textbf{0.8183}} & { \textbf{22.51}} & { \textbf{0.8004}} \\ \hline
{ }                                & { Restormer \cite{zamir2022restormer}} & { 23.50}          & { 0.7990}          & { 22.89}          & { 0.8034}          & { 22.29}          & { 0.8175}          & { 22.94}          & { 0.8059}          \\
{ }                                & { Ours(CLC)-Restormer}          & { \textbf{23.64}} & { \textbf{0.8055}} & { \textbf{23.01}} & { \textbf{0.8062}} & { \textbf{22.66}} & { \textbf{0.8209}} & { \textbf{23.15}} & { \textbf{0.8102}} \\
\rowcolor{mycolor!25} \multirow{-6}{*}{\cellcolor{white}{}} & { Ours(CSA-CLC)-Restormer}      & { \textbf{23.97}} & { \textbf{0.8102}} & { \textbf{23.34}} & { \textbf{0.8124}} & { \textbf{22.81}} & { \textbf{0.8246}} & { \textbf{23.37}} & { \textbf{0.8152}} \\ \cline{2-10} 
{ }                                & { RainRobust \cite{ba2022not}}         & { 23.46}          & { 0.7957}          & { 22.61}          & { 0.7962}          & { 22.11}          & { 0.8143}          & { 22.78}          & { 0.8013}          \\
{ }                                & { Ours-RainRobust}              & { \textbf{23.68}} & { \textbf{0.7987}} & { \textbf{22.71}} & { \textbf{0.7971}} & { \textbf{22.54}} & { \textbf{0.8163}} & { \textbf{23.02}} & { \textbf{0.8033}} \\
\rowcolor{mycolor!25} \multirow{-6}{*}{\cellcolor{white}{WeatherStream \cite{zhang2023weatherstream}}} & { Ours(CSA-CLC)-RainRobust}     & { \textbf{23.93}} & { \textbf{0.8058}} & { \textbf{23.22}} & { \textbf{0.8061}} & { \textbf{22.78}} & { \textbf{0.8237}} & { \textbf{23.38}} & { \textbf{0.8106}} \\ \hline
\end{tabular}
\end{table*}

\begin{figure*}[ht]
  \centering
  {\includegraphics[width=0.98\textwidth]{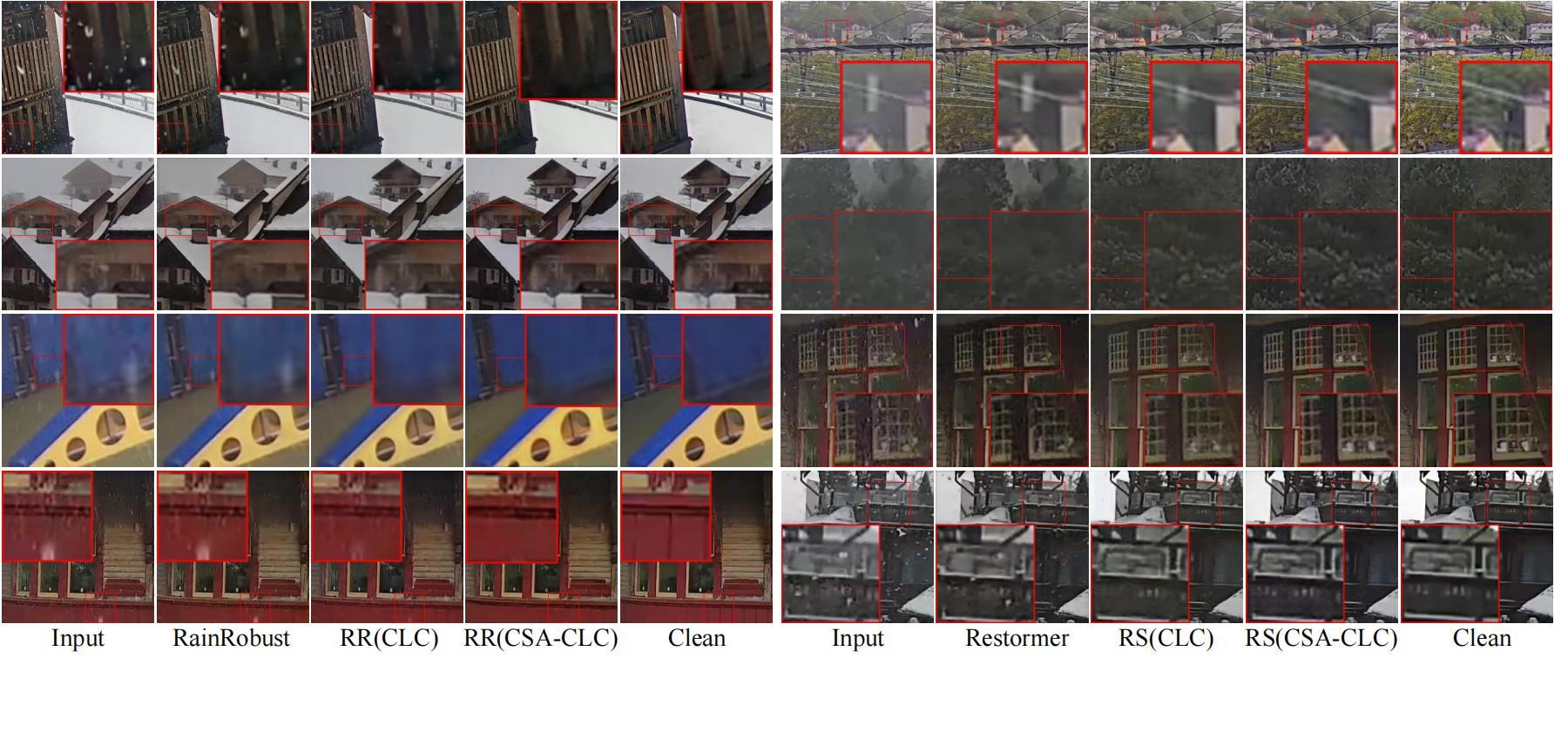}} \vspace{-3em}
  \caption{Visual comparisons of de-weathering results from our proposed pipeline with various backbone networks. Among these RR and RS represent Rain-Robust \cite{ba2022not} and Restormer \cite{zamir2022restormer} backbone, respectively.}
  \label{fig:bkbone_cpr}
\end{figure*}


To illustrate the flexible adaptability of our proposed learning framework, we utilize two backbone architectures for evaluation: the convolution-based RainRobust~\cite{ba2022not} and the transformer-based Restormer~\cite{zamir2022restormer}. Quantitative and visual results are presented in Table \ref{tab:5.4_1} and Fig. \ref{fig:bkbone_cpr} for the GT-RAIN-SNOW and WeatherStream datasets.

Across different backbone networks, our pseudo-label guided learning framework consistently demonstrates superior performance compared to training solely on original labels. Furthermore, the CSA-CLC pipeline enhances performance by implementing the Cross-frame Similarity Aggregation (CSA) module across both backbone configurations, surpassing the original CLC featured in our conference version.

\subsection{Evaluation on Pseudo-Labels} \label{sec:5.5}
\begin{figure*}[ht]
    \centering
    {\includegraphics[width=0.98\textwidth]{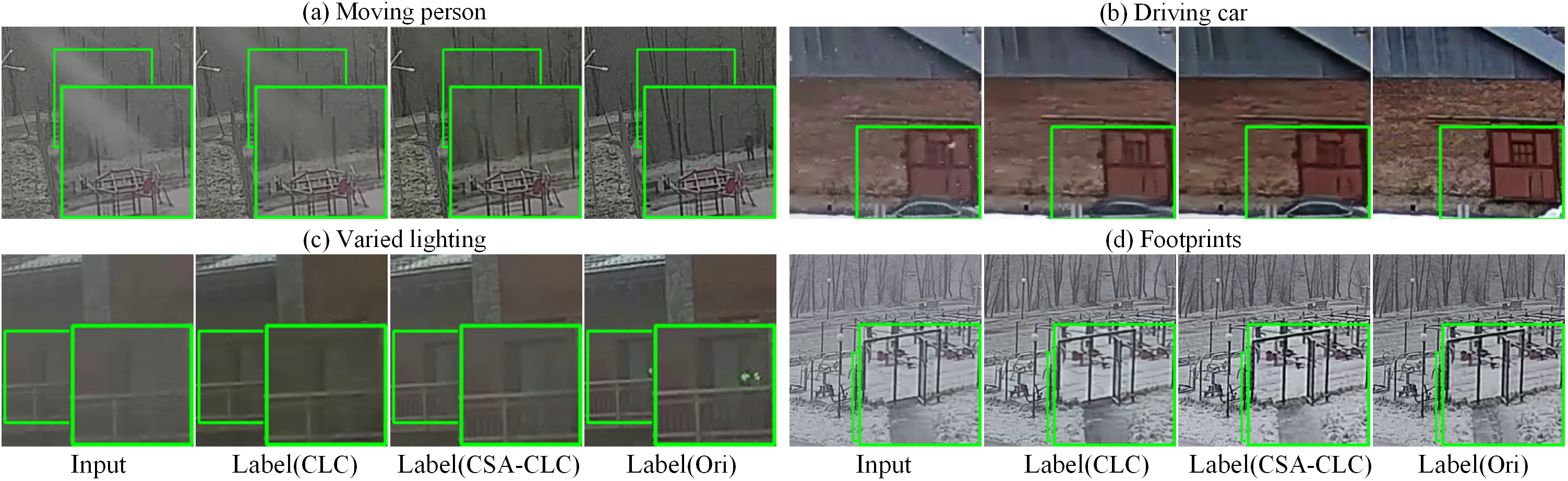}} \vspace{-0.5em}
    \caption{Visual comparisons of pseudo-labels generated by CLC and CSA-CLC with original labels. Our designed constructors produce pseudo-labels that aligned with the degraded input content, compared with original labels.}
    \label{fig:psdo_cpr}
\end{figure*}
The pseudo-labels generated through our CSA-CLC demonstrate superior consistency with degraded inputs when compared to the original labels, particularly in terms of color fidelity and object positioning. 
Fig. \ref{fig:psdo_cpr} illustrates the visual comparisons of the generated pseudo-labels. It can be observed that both techniques effectively align the pseudo-labels with the degraded inputs. Specifically, the aligned pseudo-labels in scenes (a), (c), and (d) successfully eliminate inconsistent information such as person, light reflections, and footprints present in the original labels, while scene (b) retains the driving car from degraded input. 
Furthermore, compared to CLC, the pseudo-labels generated by CSA-CLC achieve significant improvements in metric scores, as shown in Table \ref{tab:5.5_1}. Although De-W within CSA-CLC pipeline does not achieve the same level of improvement due to the limitations of single-frame input information, it nonetheless steadily improves performance thanks to high-quality pseudo labels and supplemented cross-frame aggregated features, compared with CLC-based.

\begin{table}[H]
    \centering
    \caption{Quantitative comparisons of pseudo-labels generated by CLC and CSA-CLC. \label{tab:5.5_1}}
    \resizebox{0.28\textwidth}{!}{
    \begin{tabular}{ccc}
    \hline
    Method   & PSNR$\uparrow$ & SSIM$\uparrow$  \\ \hline
    CLC  & 23.30          & 0.8282          \\
    CSA-CLC    & \textbf{26.74} & \textbf{0.8781} \\ \hline
    \end{tabular}
    }
\end{table}

\subsection{Ablation Study} 
\label{sec:5.6}
We conduct ablation studies to investigate the effectiveness of different components and settings in our method. All experiments are performed on the GT-RAIN-SNOW dataset \ref{sec:5.1}, utilizing the backbone of RainRobust for CSA-CLC. Visual comparisons can be found in supplementary material.

\subsubsection{Inconsistency-Handling Methods Comparisons} 
\label{sec:5.6.1}
\begin{table}[t]
    \centering
    \caption{Quantitative comparisons of De-W results with other inconsistency-handling methods. Baseline refers to the model trained only with single-frame original labels. \label{tab:5.6.1_1}}
    \resizebox{0.45\textwidth}{!}{
    \begin{tabular}{lll}
    \hline
    Methods                                    & PSNR$\uparrow$ & SSIM$\uparrow$  \\ \hline
    Baseline                                   & 22.01          & 0.7871          \\ \hline
    Guided Filter                              & 22.14          & 0.7959          \\
    Optical Flow                               & 22.15          & 0.7932          \\
    Misalign-Tolerate $\ell_1$                 & 22.18          & 0.7951          \\ \hline
    Guided Filter + Optical Flow               & 22.21          & 0.7935          \\
    Guided Filter + Misalign-Tolerate $\ell_1$ & 22.24          & 0.7976          \\ \hline
    Ours(CLC)                                  & \textbf{22.40} & \textbf{0.7977} \\
    Ours(CSA-CLC)                              & \textbf{22.51} & \textbf{0.8004} \\ \hline
    \end{tabular}
    }
\end{table}


Here, we compare our method with other alternative works that are capable of addressing the inconsistency between input and labels. Specifically, guided filters \cite{he2012guided} are employed for color alignment between the output and ground truth (GT) \cite{wei2020learning, wang2023benchmark}. Optical flow \cite{sun2018pwc} is utilized for spatial alignment \cite{zhang2021learning, li2023learning}, and misalignment-tolerant loss \cite{xia2023image} is used for solving misalignment. We train the De-W model by applying each method individually as well as in combination, on the same dataset, refer to Table \ref{tab:5.6.1_1}. 
The results indicate that all methods attempting to align labels outperform the baseline trained with original labels, with our approach demonstrating superior performance. Our series of methods achieve alignment through the integration of multiple frames in parameter space, particularly CSA-CLC, which performs precise matching and aggregation at a fine-grained level. However, other methods, eg., optical flow, guided filter, exhibit limited adaptability to misalignment issues at detailed levels such as lighting and texture.

\begin{table}[]
    \renewcommand{\arraystretch}{1.2}
    \centering
    \caption{Quantitative comparisons of De-weathering results from CSA-CLC pipeline under different Input Frame Numbers. \label{tab:5.6.2_1}}
    \resizebox{0.45\textwidth}{!}{
    \begin{tabular}{ccccc}
    \hline
    \multirow{2}{*}{Frame Number} & \multicolumn{2}{c}{CSA-CLC}      & \multicolumn{2}{c}{De-W} \\ \cline{2-5} 
                                  & PSNR$\uparrow$ & SSIM$\uparrow$  & PSNR$\uparrow$  & SSIM$\uparrow$  \\ \hline
    1                             & 25.32          & 0.8677          & 22.40           & 0.7986          \\
    3                             & \textbf{26.74} & 0.8781          & \textbf{22.51}  & \textbf{0.8004} \\
    5                             & 26.71          & 0.8781          & 22.48           & 0.8001          \\
    7                             & 26.73          & \textbf{0.8782} & 22.49           & 0.8002          \\ \hline
    \end{tabular}
    }
\end{table}


\subsubsection{Effect of Frame Number} 
\label{sec:5.6.2}
We investigate the performance of label constructor CSA-CLC under varying numbers of input frames (CSA-CLC performing Flexible Patch Matching within the input frame itself when the number is 1). As shown in Table \ref{tab:5.6.2_1}. 
The quality of pseudo-labels generated by CSA-CLC improves with the number of input frames, peaking at 3 frames. The De-W results also follow this trend. The self-similar complementary information from 3 frames is sufficient to capture degradation distribution. More input frames fail to stably enhance performance due to over-fitting and redundant information.
Notably, CSA-CLC with single-frame input outperforms CLC with five-frame input (Table \ref{tab:5.5_1}). there are two primary reasons for this: First, CSA-CLC incorporates original labels into its generation process, thereby providing additional information for label construction in the feature space; Second, even without auxiliary information from neighboring frames in single-frame inputs, CSA can perform Flexible Cross-frame Matching (FCM) in the current image. According to non-local self-similarity theory \cite{zontak2011internal}, CSA can identify fine-grained similar features around the query patch, thereby augmenting the available information.

\begin{table}[]
    \renewcommand{\arraystretch}{1.2}
    \centering
    \caption{Quantitative comparisons of De-W results from CSA-CLC pipeline trained under different supervision schemes. \label{tab:5.6.3_1}}
    \resizebox{0.42\textwidth}{!}{
    \begin{tabular}{cccc}
    \hline
    \multirow{2}{*}{Pseudo Label} & \multirow{2}{*}{Original Label} & \multicolumn{2}{c}{De-W} \\ \cline{3-4} 
                                  &                                 & PSNR$\uparrow$  & SSIM$\uparrow$  \\ \hline
    $\times$                      & $\checkmark$                    & 22.01           & 0.7871          \\
    $\checkmark$                  & $\times$                        & 22.42           & 0.7991          \\
    $\checkmark$                  & $\checkmark$                    & \textbf{22.51}  & \textbf{0.8004} \\ \hline
    \end{tabular}
    }
\end{table}
\subsubsection{Effect of Supervision Setting} 
\label{sec:5.6.3}
In Table \ref{tab:5.6.3_1}, we compare the performance of CSA-CLC pipeline under different label supervision settings. It is evident that they follow a consistent trend: under single-label supervision, models supervised by original labels perform worse than those using input-aligned pseudo-labels; in addition, joint supervision of both labels improves performance compared to single-label supervision. These findings prove the effectiveness of our designed joint supervision architecture, complemented by the corresponding Information Allocation Strategy (IAS), in mitigating misalignment issue in real-world de-weathering datasets, thereby ensuring high-quality supervised learning for the model.

\subsubsection{Effect of Non-Local Feature Aggregation Method} 
\label{sec:5.6.4}
\begin{table}
    \centering
    \caption{Quantitative comparisons of De-weathering results from CSA-CLC pipeline with other non-local feature aggregation methods. \label{tab:5.6.4_1}}
    \resizebox{0.45\textwidth}{!}{
    \begin{tabular}{lcccc}
    \hline
    \multirow{2}{*}{Method} & \multicolumn{2}{c}{CSA-CLC}      & \multicolumn{2}{c}{De-W}       \\ \cline{2-5} 
                            & PSNR$\uparrow$ & SSIM$\uparrow$  & PSNR$\uparrow$ & SSIM$\uparrow$  \\ \hline
    Non-local Net \cite{zhang2019residual}          & 26.45          & 0.8733          & 22.43          & 0.7993          \\
    Transformer \cite{vaswani2017attention}          & 26.57          & 0.8742          & 22.44          & 0.7995          \\
    GNN \cite{scarselli2008graph}                    & 26.62          & 0.8762          & 22.47          & 0.7996          \\ \hline
    Ours                    & \textbf{26.74} & \textbf{0.8781} & \textbf{22.51} & \textbf{0.8004} \\ \hline
    \end{tabular}
    }
\end{table}


To evaluate the differences in utilizing various feature aggregation methods during the CSA process, we fuse the matched patch sets from FCM using common non-local aggregation methods, including, Transformer \cite{vaswani2017attention}, Non-local Net \cite{wang2018non,zhang2019residual}, and GNN \cite{scarselli2008graph}. The quantitative results in Table \ref{tab:5.6.4_1} 
demonstrate that our pre-set GAT performs the best among other feature aggregation methods. The global attention mechanism of traditional Transformers tends to introduce ambiguous dependencies when addressing direction-sensitive aggregation tasks, whereas graph structures are more appropriately processed by graph networks; Non-local Net calculates aggregation relevance based on defined rules, exhibits insufficient flexibility; while the GNN, although suitable for handling directed graph data structures, computes aggregation relevance through averaging/summing, lacking real-time adaptive adjustment, and is therefore inferior to GAT.

\subsubsection{Effect of Improvement Components in CSA-CLC} \label{sec:5.6.6}
\begin{table}
\centering
\caption{Quantitative comparisons of De-weathering results after eliminating different components of CSA-CLC. \label{tab:5.6.6_1}}
\resizebox{0.44\textwidth}{!}{
\begin{tabular}{lcccc}
\hline
\multirow{2}{*}{Setting} & \multicolumn{2}{c}{CSA-CLC}      & \multicolumn{2}{c}{De-W}         \\ \cline{2-5} 
                         & PSNR$\uparrow$ & SSIM$\uparrow$  & PSNR$\uparrow$ & SSIM$\uparrow$  \\ \hline
w/o CSA                  & 24.71          & 0.8576          & 22.39          & 0.7985          \\
w/o GT input             & 24.59          & 0.8567          & 22.37          & 0.7982          \\
w/o $\mathcal{L}_{dstl}$              & 26.74          & 0.8781          & 22.46          & 0.7995          \\
Full setting             & \textbf{26.74} & \textbf{0.8781} & \textbf{22.51} & \textbf{0.8004} \\ \hline
\end{tabular}
}
\end{table}
We investigate the effectiveness of improvements by ablating components of CSA-CLC. Experiment is conducted by removing CSA, deleting original label input, and disabling distillation loss, with results presented in Table \ref{tab:5.6.6_1}.
It is evident that the absence of either CSA or original label input significantly degrades the quality of generated pseudo-labels, thereby substantially affecting the performance of De-W.
Removing the distillation loss has a relatively minor effect on De-W. This is due to the lack of multi-frame information in De-W. Distillation only enable De-W to approximate cross-frame aggregation feature distribution of CSA-CLC, which is effective to some extent, it fails to yield as significant performance improvements as directly using higher-quality pseudo-labels for supervision.
The optimal results under the full setting validate the effectiveness of our various improvements to CLC.

\begin{table}
\centering
\caption{Quantitative comparisons of De-weathering results from CSA-CLC pipeline on varying Region Padding Size ($\text{P}$). \label{tab:5.6.5_1}}
\resizebox{0.4\textwidth}{!}{
\begin{tabular}{lcccc}
\hline
\multirow{2}{*}{Padding} & \multicolumn{2}{c}{CSA-CLC}      & \multicolumn{2}{c}{De-W}       \\ \cline{2-5} 
                        & PSNR$\uparrow$ & SSIM$\uparrow$  & PSNR$\uparrow$ & SSIM$\uparrow$  \\ \hline
$\text{P}=0$            & 25.78          & 0.8697          & 22.41          & 0.7988          \\
$\text{P}=1$            & 26.53          & 0.8764          & 22.47          & 0.7997          \\
$\text{P}=3$            & \textbf{26.74} & \textbf{0.8781} & \textbf{22.51} & \textbf{0.8004} \\
$\text{P}=5$            & 26.72          & \textbf{0.8781} & 22.50          & 0.8002          \\ \hline
\end{tabular}
}
\end{table}

\subsubsection{Effect of FCM Settings in CSA-CLC} \label{sec:5.6.5}
\begin{table}
\centering
\caption{Quantitative comparisons of de-weathering results from CSA-CLC pipeline on varying selecting top-$K$ similarity patches. \label{tab:5.6.5_2}}
\resizebox{0.4\textwidth}{!}{
\begin{tabular}{lcccc}
\hline
\multirow{2}{*}{top-$K$} & \multicolumn{2}{c}{CSA-CLC}      & \multicolumn{2}{c}{De-W}       \\ \cline{2-5} 
                     & PSNR$\uparrow$ & SSIM$\uparrow$  & PSNR$\uparrow$ & SSIM$\uparrow$  \\ \hline
$K=1$         & 26.53          & 0.8758          & 22.46          & 0.7995          \\
$K=2$         & 26.62          & 0.8769          & 22.48          & 0.7997          \\
$K=3$         & \textbf{26.74} & \textbf{0.8781} & \textbf{22.51} & \textbf{0.8004} \\
$K=4$         & 26.71          & 0.8778          & 22.49          & 0.8001          \\ \hline
\end{tabular}
}
\end{table}

We conduct ablation experiments on the settings of Flexible Cross-frame Matching (FCM) in CSA-CLC, including Region Padding Size $\text{P}$ and the top-$K$ self-similar features. As shown in Table \ref{tab:5.6.5_1}, 
with the increase of $\text{P}$, FCM possesses a broader search space within adjacent frames, thereby identifying more useful complementary feature patches from adjacent frame for the query patch. When $\text{P}=3$ with a $9\times 9$ search region size, the performance reaches its peak; further increasing $\text{P}$ does not yield better results because the non-local self-similar features become sparser as the distance from the query patch increases \cite{zontak2011internal}. Consequently, an excessively large search range not only fails to enhance performance but also introduces additional computational overhead.
\ref{sec:5.6.2}.
In addition, matching more adjacent frame patches can provide richer information supplementation for query patch. While, when the number of aggregated features exceeds 3 (i.e., top-$K>3$), aggregating a surplus of patches with moderate relevance introduces redundant noise, which is not beneficial to the model, refer to Table \ref{tab:5.6.5_2}.

\section{Conclusion} 
\label{sec:6}
In this paper, we propose an innovative training paradigm for image de-weathering with non-ideal supervision, a pseudo-label guided learning framework to tackle the challenge of various inconsistencies between training pairs in the real-world dataset. 
The proposed method is composed of a de-weathering model and a Consistent Label Constructor (CLC) that is capable of producing clear pseudo-labels consistent with the degraded inputs in non-weather related content.  
Moreover, we introduce Cross-frame Similarity Aggregation (CSA) within CLC to aggregate cross-frame self-similar features from time-series inputs through Graph Attention Network, thereby enhancing the quality of the generated pseudo-labels and providing improved representation for feature distillation.
To mitigate the impact of inconsistencies on learning undesirable deformation mapping, we exploit the Information Allocation Strategy (IAS) to integrate the original ground-truth images and pseudo-labels, to facilitate the joint supervision for the de-weathering model.
Extensive experiments demonstrate that our approach outperforms other state-of-the-art techniques in handling label misaligned time-multiplexed datasets. 
Given the inevitable occurrence of imperfect label alignment in real-world de-weathering datasets, our method ensures performance while reducing the dependency on high-quality data, which is crucial for advancing the development of real-world de-weathering tasks in practical applications.

\bibliographystyle{IEEEtran}
\bibliography{ref}

\end{document}


\title{}
\maketitle

\appendix
\renewcommand{\thefigure}{\Alph{figure}}
\setcounter{figure}{0}
\section*{Sliced Wasserstein Loss} \label{apdx:A}
As illustrated in Alg.~\ref{alg:sw_loss}, the computation of the Sliced Wasserstein (SW) loss between the generated image $\mathbf{\hat{G}}$ and the reference image $\mathbf{G}$ involves several steps. Initially, we extract $2$-dimensional features from VGG19, which are subsequently transformed into $1$-dimensional representations via random linear projection. The Wasserstein distance is then determined by evaluating the element-wise $\ell_1$ distance across the sorted $1$-dimensional probability distributions of the output and target images.

\section*{Rain-Robust Loss} \label{apdx:B}
The Rain-Robust loss, draws inspiration from recent advancements in contrastive learning. A detailed description of the Rain-Robust loss is provided in Alg.~\ref{alg:rain_robust_loss}. For each image pair in the set $\{(\mathbf{I}_i, \mathbf{G}_i)\}_{i=1}^N$, we commence by extracting features using a designated feature extractor. Subsequently, we apply the InfoNCE criterion to compute the Rain-Robust loss for each positive pair. The final step involves calculating the mean Rain-Robust loss across all $N$ image pairs. During the training phase of the De-weathering Model (De-W), we input $N$ pairs of degraded-clean images from the same batch into the Rain-Robust loss framework, and designate the encoder $\mathcal{D_E}(\cdot, \Theta_\mathcal{E})$ of De-W $\mathcal{D} (\cdot, \Theta_\mathcal{D})$ as the feature extractor. $\tau$ is configured at 0.25.

\section*{Architecture of CLC} \label{apdx:C}
\begin{figure}[!ht]
    \centering
        \centering
        \includegraphics[width=0.75\linewidth]{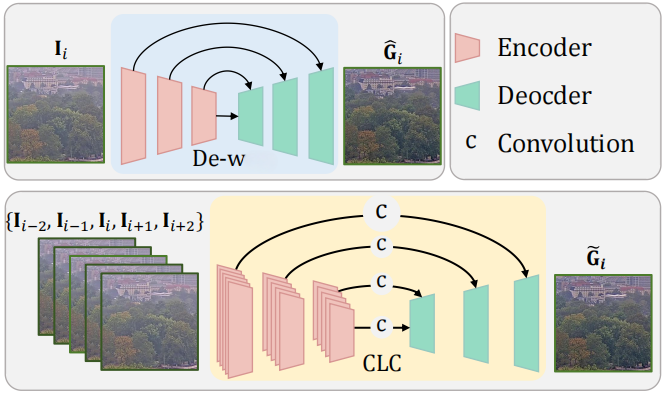}
    \caption{The Architecture of CSA-CLC and De-W.}
  \label{apdx_fig:clc_archt}
\end{figure}
Except for the additionally added modules, the architectures of CLC and CSA-CLC are essentially consistent. The architecture, depending on the experimental setting, utilizes the backbone of either Restormer or RainRobust, and both are built upon the U-Net framework, comprising an encoder and a decoder. As depicted in Fig. \ref{apdx_fig:clc_archt}, we retain the original decoder while modifying the architecture to include multiple encoders, ensuring that each input frame is processed by a distinct encoder. To facilitate the integration of multiple encoder outputs into the original decoder, we concatenate these outputs and employ convolutional layers to align their channel dimensions. The architecture of De-w is consistent with that of the pseudo-label constructor, with the difference being the modification from multi-frame input to single-frame input.

\begin{algorithm}[!ht]
\caption{Pseudo code of SW loss}
\begin{algorithmic}[1]
    \REQUIRE $\mathbf{\hat{G}}$: output image, \\
             $\mathbf{G}$: target image, \\
             $\mathbf{M}\in\mathbb{R}^{C' \times C}$: random projection matrix.
    \STATE
        obtain VGG19 features: $\mathbf{U}=\text{VGG}(\mathbf{\hat{G}})$, $\mathbf{V}=\text{VGG}(\mathbf{G})$, $\mathbf{U}$ and $\mathbf{V}$ are both in $\mathbb{R}^{C \times H \times W}$.
    \STATE 
        reshape the feature $\mathbf{U}$ and $\mathbf{V}$ to $\mathbf{U_r}\in\mathbb{R}^{C \times HW}$ and $\mathbf{V_r}\in\mathbb{R}^{C \times HW}$, respectively.
    \STATE
        project the feature $\mathbf{U}$ and $\mathbf{V}$ to $\mathbb{R}^{C' \times HW}$ using $\mathbf{M}$: $\mathbf{U_p}=\mathbf{M}\mathbf{U_f}$, $\mathbf{V_p}=\mathbf{M}\mathbf{V_f}$.
    \STATE
        sort $\mathbf{U_p}$ and $\mathbf{V}_p$: $\mathbf{U_s}=\text{Sort}(\mathbf{U_p}, \text{dim}=1)$, $\mathbf{V_s}=\text{Sort}(\mathbf{V_p}, \text{dim}=1)$.
    \RETURN $\Vert\mathbf{U_s}-\mathbf{V_s}\Vert_1$.
\end{algorithmic}
\label{alg:sw_loss}
\end{algorithm}

\begin{algorithm}[!ht]
\caption{Pseudo code of Rain-Robust loss}
\begin{algorithmic}[1]
    \REQUIRE $\{(\mathbf{I}_i, \mathbf{G}_i)\}_{i=1}^N$: $N$ degraded-clean image pairs from different scenes,\\
             $\mathcal{D_E}(\cdot, \Theta_\mathcal{E})$: encoder of the current De-W,\\
             $\tau$: temperature coefficient.
    \FOR {{$i$ from $1$ to $N$}}
        \STATE
            $\mathbf{U}_i=\mathcal{D_E}(\mathbf{I}_i, \Theta_\mathcal{E})$ and $\mathbf{V}_i=\mathcal{D_E}(\mathbf{G}_i, \Theta_\mathcal{E})$,
    \ENDFOR
    \FOR {{$i$ from $1$ to $N$}}
        \STATE \hspace{-1.5em}
        $\mathcal{L}_{\mathbf{VU}_i} = -\text{log}\frac{\text{exp}(\text{sim}_{\text{cos}}(\mathbf{U}_i, \mathbf{V}_i)/\tau)}{\Sigma_{j=1, j\neq i}^N(\text{exp}(\text{sim}_{\text{cos}}(\mathbf{U}_i, \mathbf{U}_j)/\tau) + \text{exp}(\text{sim}_{\text{cos}}(\mathbf{U}_i, \mathbf{V}_j)/\tau))}$\\
        \hspace{-1.5em}
        $\mathcal{L}_{\mathbf{UV}_i} = -\text{log}\frac{\text{exp}(\text{sim}_{\text{cos}}(\mathbf{V}_i, \mathbf{U}_i)/\tau)}{\Sigma_{j=1, j\neq i}^N(\text{exp}(\text{sim}_{\text{cos}}(\mathbf{V}_i, \mathbf{U}_j)/\tau) + \text{exp}(\text{sim}_{\text{cos}}(\mathbf{V}_i, \mathbf{V}_j)/\tau))}$
    \ENDFOR
        \RETURN $\frac{\Sigma_{i=1}^N(\mathcal{L}_{\mathbf{VU}_i} + \mathcal{L}_{\mathbf{UV}_i})}{N}$.
\end{algorithmic}
\label{alg:rain_robust_loss}
\end{algorithm}

\section*{Visual Comparisons of Ablation Study} \label{apdx:E}
\begin{figure*}[ht]
  \centering
  \captionsetup[subfloat]{labelformat=empty}
  \subfloat[\footnotesize (a) Input]
  {\includegraphics[width=0.14\textwidth]{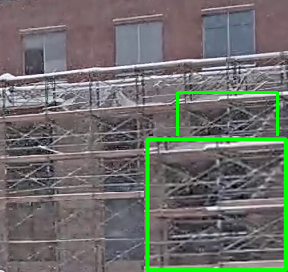}}
  \subfloat[\footnotesize (b) Baseline]
  {\includegraphics[width=0.14\textwidth]{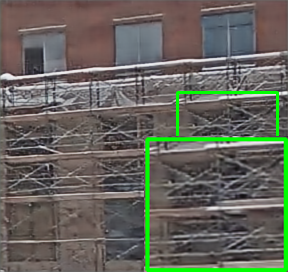}}
  \subfloat[\footnotesize (c) GF+OF]
  {\includegraphics[width=0.14\textwidth]{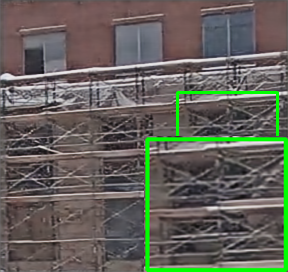}}
  \subfloat[\footnotesize (d) GF+MT $\ell_1$]
  {\includegraphics[width=0.14\textwidth]{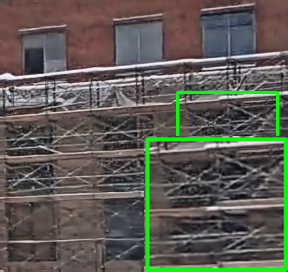}}
  \subfloat[\footnotesize (e) Ours(CLC)]
  {\includegraphics[width=0.14\textwidth]{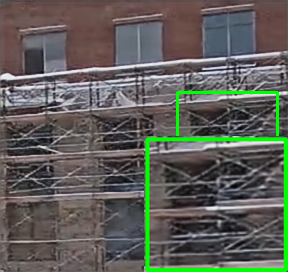}}
  \subfloat[\footnotesize (f) Ours(CSA-CLC)]
  {\includegraphics[width=0.14\textwidth]{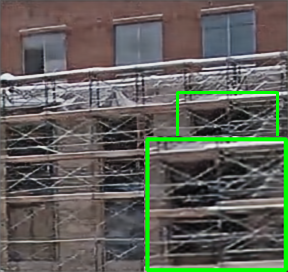}}
  \subfloat[\footnotesize (g) Clean]
  {\includegraphics[width=0.14\textwidth]{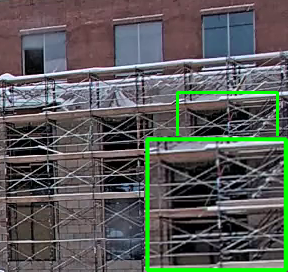}}
  \caption{Visual comparisons of De-W results under other inconsistency-handling methods. Among these GF, OF, and MT represent Guided Filter, Optical Flow, and Misalign-Tolerate $\ell_1$, respectively.}
  \label{fig:abl_align_cpr}
\end{figure*}

\begin{figure*}[ht]
  \centering
  \captionsetup[subfloat]{labelformat=empty}
  \subfloat[\footnotesize Input]
  {\includegraphics[width=0.15\textwidth]{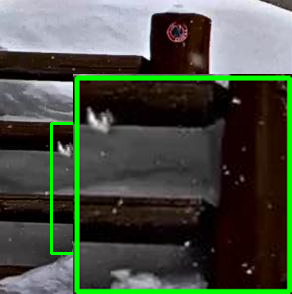}}\vspace{0.1em}
  \subfloat[\footnotesize (e) $\text{Frame}=1$]
  {\includegraphics[width=0.15\textwidth]{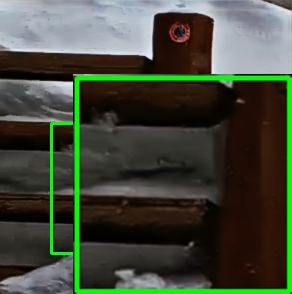}}\vspace{0.1em}
  \subfloat[\footnotesize (f) $\text{Frame}=3$]
  {\includegraphics[width=0.15\textwidth]{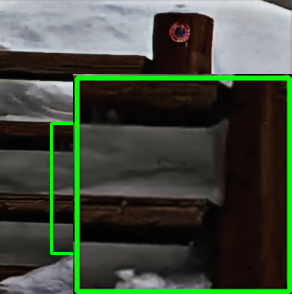}}\vspace{0.1em}
  \subfloat[\footnotesize (g) $\text{Frame}=5$]
  {\includegraphics[width=0.15\textwidth]{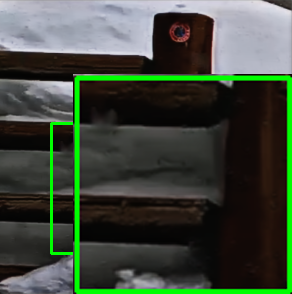}}\vspace{0.1em}
  \subfloat[\footnotesize (h) $\text{Frame}=7$]
  {\includegraphics[width=0.15\textwidth]{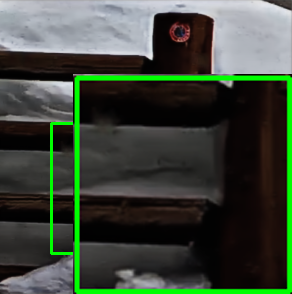}}\vspace{0.1em}
  \subfloat[\footnotesize Clean]
  {\includegraphics[width=0.15\textwidth]{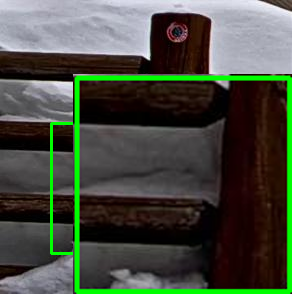}}\vspace{0.1em}
  \caption{Visual comparisons of De-W results from CSA-CLC pipeline under different Input Frame Numbers.}
  \label{fig:abl_N_cpr}
\end{figure*}

\begin{figure*}[ht]
  \centering
  \captionsetup[subfloat]{labelformat=empty}
  \subfloat[\footnotesize (a) Input]
  {\includegraphics[width=0.13\textwidth]{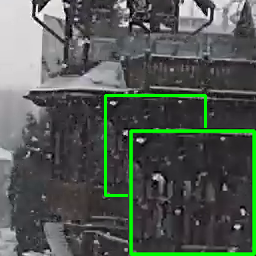}}\vspace{0.1em}
  \subfloat[\footnotesize (d) Non-local Net]
  {\includegraphics[width=0.13\textwidth]{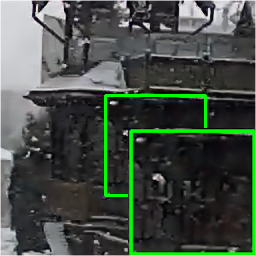}}\vspace{0.1em}
  \subfloat[\footnotesize (c) Transformer]
  {\includegraphics[width=0.13\textwidth]{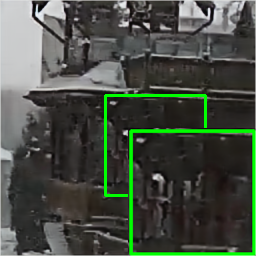}}\vspace{0.1em}
  \subfloat[\footnotesize (e) GNN]
  {\includegraphics[width=0.13\textwidth]{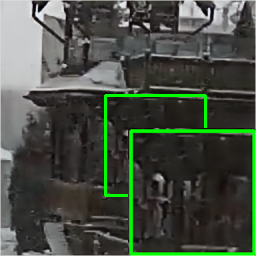}}\vspace{0.1em}
  \subfloat[\footnotesize (f) Ours(GAT)]
  {\includegraphics[width=0.13\textwidth]{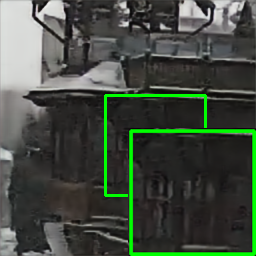}}\vspace{0.1em}
  \subfloat[\footnotesize (g) Clean]
  {\includegraphics[width=0.13\textwidth]{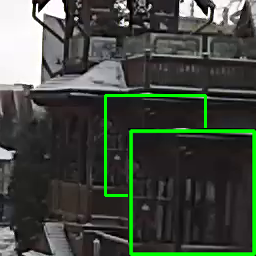}}
  \caption{Visual comparisons of De-W results from CSA-CLC pipeline with other non-local feature aggregation methods.}
  \label{fig:abl_agg_cpr}
\end{figure*}

\begin{figure*}[ht]
  \centering
  \captionsetup[subfloat]{labelformat=empty}
  \subfloat[\footnotesize (a) Input]
  {\includegraphics[width=0.13\textwidth]{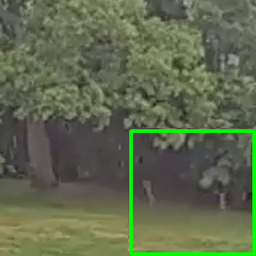}}\vspace{0.1em}
  \subfloat[\footnotesize (b) $\text{P}=0$]
  {\includegraphics[width=0.13\textwidth]{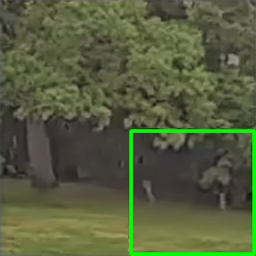}}\vspace{0.1em}
  \subfloat[\footnotesize (c) $\text{P}=1$]
  {\includegraphics[width=0.13\textwidth]{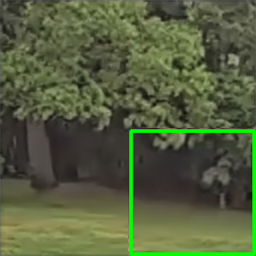}}\vspace{-1em}
  \subfloat[\footnotesize (d) $\text{P}=3$]
  {\includegraphics[width=0.13\textwidth]{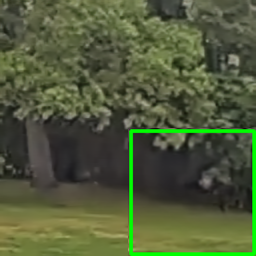}}\vspace{0.1em}
  \subfloat[\footnotesize (e) $\text{P}=5$]
  {\includegraphics[width=0.13\textwidth]{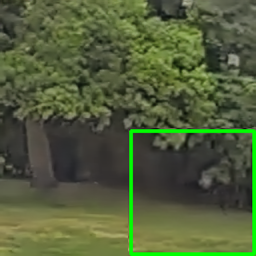}}\vspace{0.1em}
  \subfloat[\footnotesize (f) Clean]
  {\includegraphics[width=0.13\textwidth]{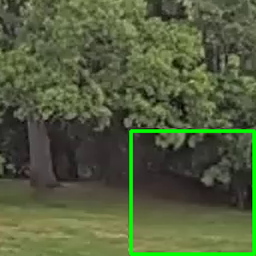}}
  \caption{Visual comparisons of De-W results from CSA-CLC pipeline on varying Region Padding Size ($\text{P}$).}
  \label{fig:abl_P_cpr}
\end{figure*}

\begin{figure*}[ht]
  \centering
  \captionsetup[subfloat]{labelformat=empty}
  \subfloat[\footnotesize (a) Input]
  {\includegraphics[width=0.13\textwidth]{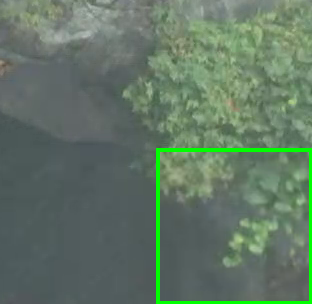}}\vspace{0.1em}
  \subfloat[\footnotesize (b) $K=1$]
  {\includegraphics[width=0.13\textwidth]{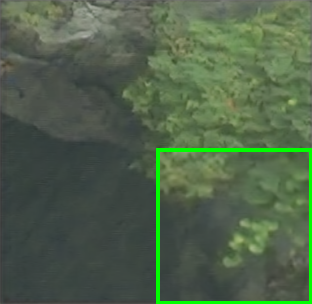}}\vspace{0.1em}
  \subfloat[\footnotesize (c) $K=2$]
  {\includegraphics[width=0.13\textwidth]{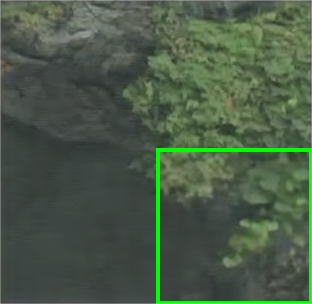}}\vspace{-1em}
  \subfloat[\footnotesize (d) $K=3$]
  {\includegraphics[width=0.13\textwidth]{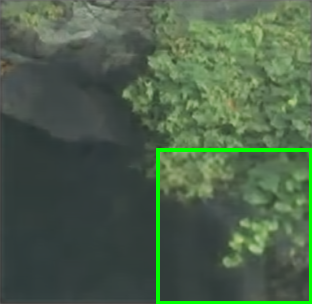}}\vspace{0.1em}
  \subfloat[\footnotesize (e) $K=4$]
  {\includegraphics[width=0.13\textwidth]{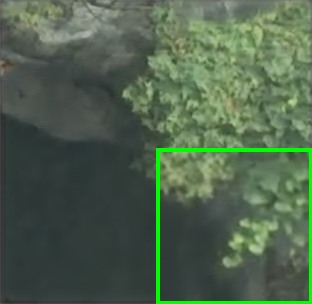}}\vspace{0.1em}
  \subfloat[\footnotesize (f) Clean]
  {\includegraphics[width=0.13\textwidth]{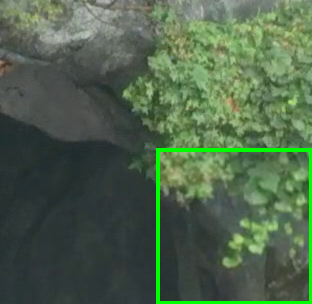}}
  \caption{Visual comparisons of De-W results from CSA-CLC pipeline on varying selecting top-$K$ similarity patches.}
  \label{fig:abl_topK_cpr}
\end{figure*}

This section presents several visual comparisons from the ablation study. Figure \ref{fig:abl_align_cpr} illustrates the visual comparison between our method and other label non-consistent alignment methods (Sec. IV-F1). Our method can effectively restore weather degradation while preserving better structural information.
Figure \ref{fig:abl_N_cpr} is the visual results of different input frame numbers (Sec. IV-F2). It can be observed that a sufficiently good snowflake removal effect can be achieved with only three input frames.
The visual comparison of various non-local feature aggregation methods in Cross-Frame Similarity Aggregation (CSA) is referenced in Figure \ref{fig:abl_agg_cpr} (Sec. IV-F4). indicating our aggregation strategy excels in snowflake removal.
Figures \ref{fig:abl_P_cpr} and \ref{fig:abl_topK_cpr} display the visual comparison under different parameter settings in Flexible Cross-Frame Matching (FCM) in Sec. IV-F5. The performance reaches a ceiling when Region Padding Size is 3 and top-$K$=3, and increasing them does not further enhance the performance but adds to the computational overhead.

